\documentclass[twocolumn,showpacs,amsmath,amssymb,nofootinbib,floatfix,aps,prd,10pt,eqsecnum]{revtex4-1}
\usepackage{dcolumn}
\usepackage{bm}
\usepackage{graphicx}
\usepackage{color}
\usepackage{hyperref}
\usepackage{amsthm}
\usepackage{fancyvrb}
 
\newcommand{\be}{\begin{equation}}
\newcommand{\ee}{\end{equation}}
\newcommand{\ba}{\begin{eqnarray}}
\newcommand{\ea}{\end{eqnarray}}
\newcommand{\non}{\nonumber}
\newcommand{\n}[1]{\label{#1}}
\newcommand{\eq}[1]{Eq.~(\ref{#1})}

\newcommand{\hhh}{\, ,\hspace{0.5cm}}
\newcommand{\BM}[1]{{\mbox{\boldmath $#1$}}}
\newcommand{\ind}[1]{\mbox{\tiny #1}}

\begin{document}

\title{Template banks based on $\mathbb{Z}^n$ and $A_n^*$ lattices}
\date{\today}
\author{Bruce Allen}
\email{bruce.allen@aei.mpg.de}
\author{Andrey A. Shoom}
\email{andrey.shoom@aei.mpg.de}
\affiliation{Max Planck Institute for Gravitational Physics (Albert Einstein Institute), Leibniz Universität Hannover, Callinstr. 38, D-30167, Hannover, Germany}


\begin{abstract}
  \noindent
  Matched filtering is a traditional method used to search a data
  stream for signals.  If the source (and hence its $n$ parameters)
  are unknown, many filters must be employed.  These form a grid in
  the $n$-dimensional parameter space, known as a template bank.  It
  is often convenient to construct these grids as a lattice.  Here, we
  examine some of the properties of these template banks for
  $\mathbb{Z}^n$ and $A_n^*$ lattices.  In particular, we focus on the
  distribution of the mismatch function, both in the traditional
  quadratic approximation and in the recently-proposed spherical
  approximation.  The fraction of signals which are lost is determined
  by the even moments of this distribution, which we calculate.  Many
  of these quantities we examine have a simple and well-defined
  $n\to\infty$ limit, which often gives an accurate estimate even for
  small $n$. Our main conclusions are the following: (i) a fairly
  effective template-based search can be constructed at mismatch
  values that are shockingly high in the quadratic approximation; (ii)
  the minor advantage offered by an $A_n^*$ template bank (compared to
  $\mathbb{Z}^n$) at small template separation becomes even less
  significant at large mismatch.  So there is little motivation for
  using template banks based on the $A_n^*$ lattice.
\end{abstract}

\maketitle

\section{Introduction}

Matched filtering is a standard technique \cite{WZ,Hels} used to
search for weak gravitational-wave signals from the binary inspiral of
black holes and/or neutron stars.  This compares the data (suitably
weighted in frequency space) to a template of the expected waveform
\cite{Schutz:1989yw,Sathyaprakash:1991mt,Schutz_Book,Cutler:1992tc,Sathyaprakash:1994nj,Cutler:1994ys,Dhurandhar:1992mw,Dhurandhar:1994mi,Balasubramanian:1994uy,Balasubramanian:1995bm,Owen:1998dk,FINDCHIRP}. Matched
filtering is also used to search for weak electromagnetic (radio and
gamma-ray) \cite{Nieder} and gravitational-wave signals from
rapidly rotating neutron stars (pulsars) \cite{cartel} and has many
other applications across a broad range of fields and topics.

Because these searches are typically looking for new events and/or
unknown sources, the parameters of the signals are not known.  Some
examples of these parameters include sky position, mass, and spin or
chirp frequency. Thus, a collection of templates must be employed.  The
grid of these templates in parameter space is generally referred to as
a ``template bank''.

If the parameter space is low-dimensional and the parameter-space
volume is not too large, one can simply ``overcover'' the space,
putting many redundant templates close together.  However, if the
parameter-space dimension and/or volume is large, this quickly becomes
(computationally speaking) very expensive.  On the other hand, if the
templates are spaced too far apart, then it's possible that
some signals be missed, because there was no template
in the bank which matched the waveform well enough.  Thus, a
compromise must be reached: enough templates must be employed that
signals are not lost, but the number of templates must not be so large
that the computing cost explodes. For some searches (e.g., for
continuous gravitational waves from neutron stars in binary systems)
the computing cost is so high that it constrains the search
sensitivity.

The problem of how to place templates in parameter space is
well studied. There are many ways to construct template banks.  For
example, one can simply place the templates at random
\cite{MessengerPrixPapa}, with a high enough density that most signals
are likely to lie near enough to a template.  Or one can improve this
by removing redundant templates which are ``too close'' to neighboring
ones, and adding more templates at random, if required
\cite{AllenHarrySathya}.  Alternatively, one can build a template bank
as a regular lattice in parameter space.  Two examples of such
lattices are the $\mathbb{Z}^n$ and $A_n^*$ lattices.  The first of these
is just the Cartesian product of equally spaced grids in all
dimensions, and the second is the $n$-dimensional generalization of
the two-dimensional hexagonal lattice and the three-dimensional
face-centered cubic (fcc) lattice.

One way to characterize a template bank is via the mismatch function
$m$. This is a function on parameter space, which quantifies how much
signal-to-noise ratio (SNR) is lost because of the discreteness of the
template bank. Its value at any point is the fractional difference
between the squared SNR obtained for a signal with those parameters,
and the squared SNR that would have been obtained had a template been
located at that point. Thus, $m$ vanishes at the locations of the
templates, and is largest ``halfway in-between'' two templates.
In a recent paper, we show how the fraction of lost signals is related
to the average of $m$ and functions of $m$ \cite{NewAllenPaper}.

When the templates are close together, and the mismatch is small, $m$
can be expressed as a positive-definite quadratic form and thought of
as the squared distance between the parameter-space point and the closest
template.  Thus, in this approximation,
$m\approx g_{ab}\Delta\lambda^a\Delta\lambda^b$, where $g_{ab}$ is the metric
on the parameter space and $\Delta\lambda^a$ is the coordinate separation between the two points (see,
e.g., \cite{Owen,Owen:1998dk}).  Here, we call this the quadratic
approximation to the mismatch, and write it as $m=r^2$.  When the
templates are less-closely spaced, a better approximation to the
mismatch is the ``spherical" ansatz, $m\approx \sin^2 r =
\sin^2(\sqrt{g_{ab}\Delta\lambda^a\Delta\lambda^b})$, recently
introduced in \cite{AllenSpherical}.

If the mismatch is small, then the bank which minimizes the average
second-moment of $r^2$ loses the smallest fraction of signals
\cite{NewAllenPaper}.  If the bank is a lattice, this is called the
``optimal quantizer'' \cite{Conway}.  This paper extends those results
to large mismatch, by exploiting the spherical ansatz
\cite{AllenSpherical}, and carrying out an explicit calculation for
template banks constructed from the $\mathbb{Z}^n$ and $A_n^*$
lattices.

Our paper is organized as follows. In Sec.~\ref{s:lattices} we
describe the $n$-dimensional lattices $\mathbb{Z}^n$ and $A_n^*$, and
derive their key properties.  In Sec.~\ref{s:lossfraction} we
calculate the fraction of lost detections in the quadratic and
spherical approximations for these lattices for $2$- and
$3$-dimensional source distribution.  This fraction of lost signals
may be thought of as the ``inefficiency'' or ``loss fraction'' of the
lattice.  In Sec.~\ref{s:largen} we evaluate the loss fraction as the
parameter space dimension $n \to \infty$.  This gives simple analytic
expressions; in some cases the approach is fast enough that these are
good approximations even in finite numbers of dimensions.  In
Sec.~\ref{s:compare} we compare the loss fraction of $\mathbb{Z}^n$
and $A_n^*$ at fixed computing cost.  Finally, in
Sec.~\ref{s:mismatch}, we examine the distribution function of the
squared radius $r^2$, and its properties for the $\mathbb{Z}^n$ and
$A_n^*$ lattices.  This is followed by a short Conclusion.

Our results only depend on the even-order moments of the Wigner-Seitz (WS)
cells of the lattices, which we denote by $\langle r^{2m} \rangle$.
Appendix~\ref{s:zlatticemoments} contains a calculation of these
moments for the $\mathbb{Z}^n$ lattice, and
Appendix~\ref{s:AnStarMoments} contains the corresponding calculation
for the $A_n^*$ lattice.

\section{\label{s:lattices} The $\mathbb{Z}^n$ and $A_n^*$ lattices}

\begin{figure*}[htb]
\begin{center}
\ba
&&\hspace{0cm}\includegraphics[width=4cm]{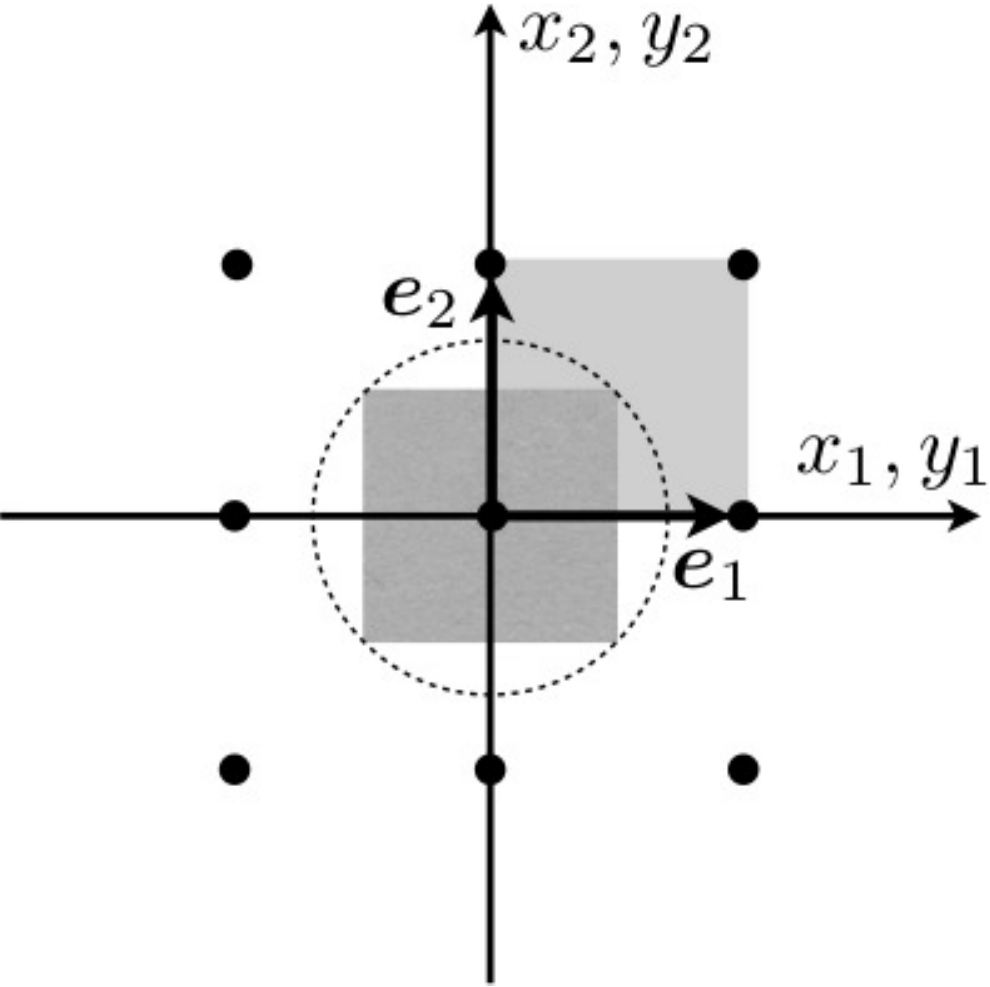}
\hspace{3cm}\includegraphics[width=4cm]{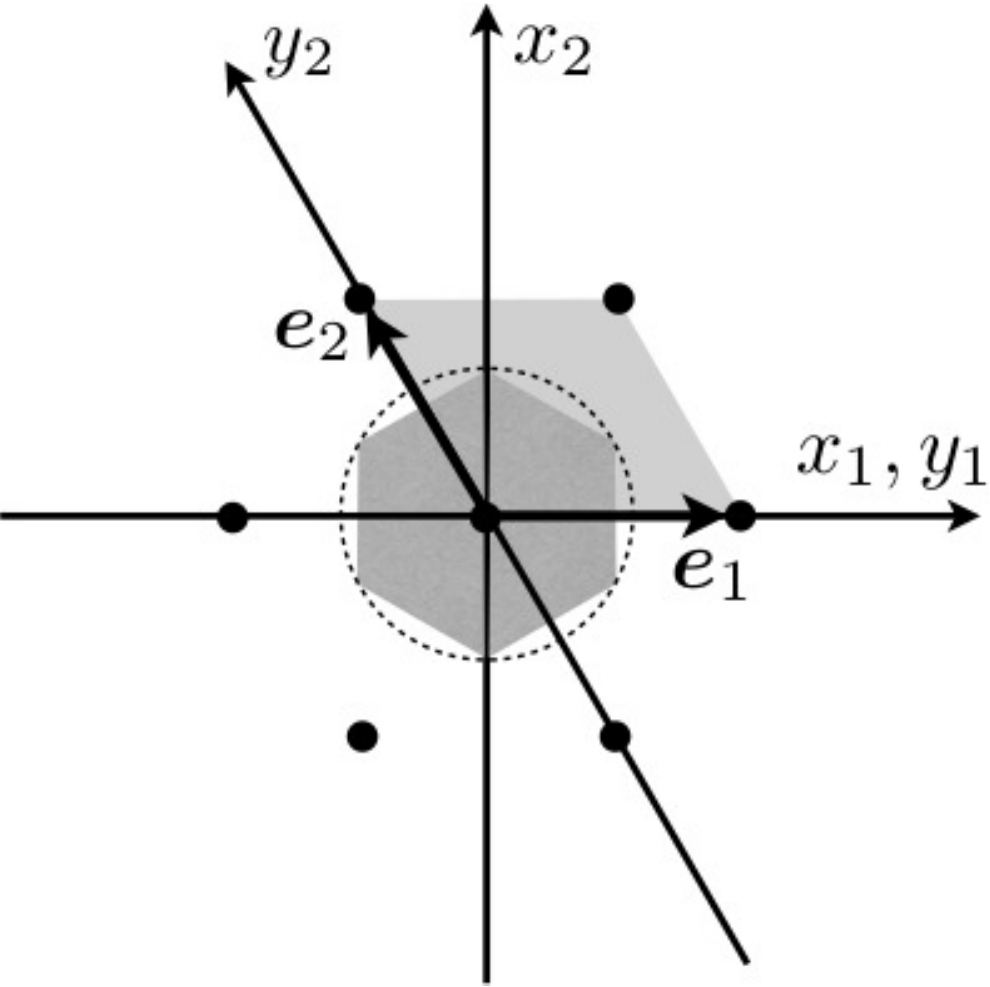}\non\\
&&\hspace{1.75cm}({\bf a})\hspace{6.6cm}({\bf b})\non
\ea
\caption{Two-dimensional lattices $(\ell=1)$: ({\bf a}) the $\mathbb{Z}^2$ square
  lattice, ({\bf b}) the $A_2^*$ hexagonal lattice. The fundamental
  polytropes are shown in light grey, the WS cells are shown
  in dark grey and inscribed by the dashed circles of the covering
  radius.  For general $n$, the basis vectors $\BM{e}_i$ for $A_n^*$ define
  vertices of an equilateral $n$-simplex (see text after \eq{2.8}).
}\label{f1}
\end{center}
\end{figure*}

The lattices $\mathbb{Z}^n$ and the $A_n^*$ are an infinite collection
of regularly-spaced points in Cartesian space $\mathbb{R}^n$.
We use $\BM{x}$ to denote a point in $\mathbb{R}^n$ with the normal
Euclidean norm ${| \BM{x} | }^2 = \BM{x} \cdot \BM{x} $, where the dot
denotes the standard dot product.  The lattices are generated by a set
of $n$ basis vectors $\BM{e}_i \in \mathbb{R}^n$, for $i \in 1,\cdots,
n$, which for these particular lattices are normalized so that
$\BM{e}_i \cdot \BM{e}_i =1 $.  Two-dimensional representatives of
these lattices are illustrated in Fig.~\ref{f1}.

To characterize the geometry of the lattice, we shall use $x_i$ to
denote Cartesian coordinates and $y_i$ to denote lattice
coordinates. Accordingly, the lattice consists of all points $\BM{x} =
y_i\BM{e}_i$ in $\mathbb{R}^n$, such that $y_i=c_i \ell$, where
$c_i\in\mathbb{Z}$ are integers and $\ell$ is the lattice
spacing. These points are called lattice {\em vertices}. From here
forward, we shall use the ``summation convention'' that repeated
indices are summed.

The squared distance $r^2$ between points $\BM{x}_A$ and $\BM{x}_B$
with lattice coordinates $y_{Ai}$ and $y_{Bi}$ is then
\be\n{2.1}
r^2 =   (\BM{x}_A - \BM{x}_B) \cdot (\BM{x}_A - \BM{x}_B)  = \Delta y_i \Delta y_j\BM{e}_i \cdot\BM{e}_j = g_{ij} \Delta y_i \Delta y_j\,,
\ee
where $\Delta y_i =  y_{Ai} - y_{Bi}$ are the lattice coordinate separations and  
the (flat) metric is $g_{ij} =  \BM{e}_i \cdot \BM{e}_j$.

The region of $\mathbb{R}^n$ for which the coordinates $y_i$'s are such that
$|y_i|\in[0,\ell]$, is called a ``Fundamental Polytope'' or FP. The FP has
$2^n$ vertices, which are neighboring lattice points.  The region of
$\mathbb{R}^n$ which is closer [in the sense of the coordinate
  distance \eq{2.1}] to a given lattice point than to any other
lattice point is called the ``Wigner-Seitz cell'' (WS) of that lattice
point.  We also denote the Wigner-Seitz cell of the origin $y_i=0$ by WS
(see Fig.~\ref{f1}).  The distance from the origin to the most distant
point of the WS is called the covering radius or WS radius $R$; it is
the radius of the smallest sphere about the origin which encloses
every point of the WS.

We can compute the $n$-volume of the FP and the WS as follows. Since
all FP are equivalent, we concentrate on the fundamental FP defined by
lattice coordinate values $y_i \in [0,\ell]$. The volume of the FP is
\be\n{2.2} 
V_{\ind{FP}} = \int_0^\ell dy_1 \cdots \int_0^\ell dy_n\sqrt{g}\,, 
\ee 
where $g=\det(g_{ij})$. The $n$-volume of the WS, $V_{\ind{WS}}$, is
equal to that of the FP, because (if the WS is copied around all
lattice points) they overlap only on the boundaries (a set of measure
zero), are in one-to-one correspondence, and cover all of space.

The fundamental FP is contractible to the origin, in the sense that if
a point $\BM{x} \in \mathbb{R}^n$ lies inside it, then so does the
point $\lambda\BM{x}$ for $\lambda \in [0,1)$.  Because it is defined
via a linear construction, it can also be contracted to any other
vertex, so it is convex and contractible in any direction. By
construction the FP is bounded by a set of $(n-1)$-dimensional planes.

In similar fashion, the boundary of the WS is defined by a set of
$(n-1)$-dimensional planes that lie halfway between the origin and the
surrounding lattice points.  We can compute the covering radius $R$ of
the WS centered at the origin by considering the subset of those
planes which lie in the FP, i.e. which lie halfway between the origin
and the remaining $2^n-1$ FP lattice points, and finding the point of
intersection most distant from the origin.

To characterize the efficiency of the space covering, one defines the
thickness $\Theta$ as the average number of covering spheres that contain
a point of the space. This is equal to the ratio of volume of an
$n$-dimensional ball enclosed by one of the spheres, to
the volume $V_{\ind{FP}}=V_{\ind{WS}}$ \cite{Conway},
\be
\Theta=\frac{V(B_n(R))}{V_{\ind{WS}}}.
\ee
Here, $V(B_n(R))=\pi^{n/2}R^n/\Gamma(1+n/2)$ is the volume of an
$n$-ball $B_n$ of radius $R$.  From the definition it follows that
$\Theta\geq1$.  Smaller values of $\Theta$ indicate less overlap among
the balls, i.e. a more efficient covering.\footnote{Another quantity
  used in the literature is the normalized thickness (or center
  density) $\theta=R^n/V_{\ind{WS}}$.}

In the following subsections we compute the quantities defined above
for the $\mathbb{Z}^n$ and $A_n^*$ lattices. We will use these quantities 
in calculating the statistical properties of
functions of the distance, such as the mismatch, for both the lattices
and to compare the derived results.

\subsection{The $\mathbb{Z}^n$ lattice}

The $\mathbb{Z}^n$ lattice  (see, e.g., \cite{Conway}) is generated by orthonormal basis vectors
\be\n{2.3}
\BM{e}_i \cdot \BM{e}_j = \delta_{ij}\,,
\ee
where $\delta_{ij}$ is the Kronecker delta, i.e., the metric $g_{ij}$ is the identity matrix. Thus,
if the basis vectors are taken as the 
standard coordinate basis, then
the lattice coordinates $y_i$ are just the normal Cartesian coordinates and $x_i=y_i$.  
The distance function \eq{2.1} is
\be\n{2.4}
r^2=\sum_{i=1}^n \Delta y_i^2\,,
\ee
and, according to \eq{2.2}, the $n$-volume of the FP is
\be\n{2.5}
V_{\ind{FP}} = \ell^n\,.
\ee 
According to \eq{2.1}, the largest distance between any pair of
vertices in the FP is $r^2=n\ell^2$. It is also the largest distance
from the origin to a point within the FP.

To find the boundary of the WS centered at the origin, we begin by
finding the equations of the planes that lie halfway between the origin
and the nearest lattice points at distance $\ell$ from the origin.
(The other potential bounding planes are irrelevant because they lie
outside.)  There are $2n$ of these nearest lattice points.  They have
coordinates $(0,\cdots, 0, \pm \ell, 0,\cdots, 0)$, where $\ell$ is
located in the $j$th position and the remaining $n-1$ coordinates
vanish.  Using the distance function \eq{2.4} we find that the
coordinates in the $(n-1)$-dimensional boundary planes satisfy the
equation
\be\non 
\sum_{i=1}^n y_i^2 =  (y_j \mp \ell)^2 + \sum_{\substack{i=1\\(i\ne j)}}^ny_i^2  = \ell^2 \mp 2 \ell y_j + \sum_{i=1}^n y_i^2 \,.
\ee
Thus the planes bounding the WS satisfy
\be\n{2.6}
y_j=\pm \ell/2\,.
\ee 
There are $2 n$ such planes, since $j=1,\cdots, n$.   These define an $n$-cube which is identical to the FP but shifted by $-\ell/2$ along each coordinate axis, so that its center is at the origin. Note that the result \eq{2.6}
follows directly from the lattice geometry.

The WS radius $R$ is easily computed.
The point of mutual intersection of the $n$ planes with $y_j>0$
defines a vertex of the WS.  All of the $2^n$ vertices of the WS
(defined by intersecting each of the possible planes, one for each
coordinate, $n$ in total) are at the same distance $R$ from the origin.
Hence, the WS covering radius $R$ is the distance of that WS vertex
from the origin. Using the expression \eq{2.4} gives
\be\n{2.7}
R^2=\frac{1}{4}n\ell^2\,
\ee
for the covering radius of the $\mathbb{Z}^n$ lattice.  The $n$-volume
of the FP and of the WS can be expressed in terms of $R$, as
\be\n{volZ}
V_{\ind{FP}}=V^{\mathbb{Z}^n}_{\ind{WS}} = 2^n n^{-n/2} R^n.
\ee
Later, we will compare the properties of different lattices at fixed $V_{WS}$.

\subsection{The $A_n^*$ lattice}

The $A_n^*$ lattice is a classical root lattice, whose attractions
have been discussed in detail by \cite{Prix_2007}.  For $n\le 17$ it is
either the thinnest classical root lattice, or close to the thinnest
one.  (Note however that thinner non-classical lattices have been
constructed numerically, by semidefinite optimization in the space of
lattices.  The current record-holders are listed in Table~2 of
\cite{Sikiric_2008}.)

The $A_n^*$ lattice is
generated by basis vectors chosen to satisfy (see, e.g., \cite{Conway,RB})
\be\n{2.8}
\BM{e}_i \cdot \BM{e}_j =
\begin{cases}
  1    & \text{for } i=j\\
  -1/n & \text{for } i \ne j.
  \end{cases}
\ee
The vectors $\BM{e}_i$ are easily visualized: they point from the origin to $n$ of
the $n+1$ vertices of an equilateral $n$-simplex. (The unit vector from the origin to the final vertex of
the simplex is $-\BM{e}_1 - \cdots -\BM{e}_n$,
which implies that the center of the simplex lies at the origin of
coordinates.)

For this lattice, the distance function \eq{2.1} is
\ba\n{2.9}
r^2&=&\sum_{i=1}^n\Delta y_i^2-\frac{1}{n}\sum_{\substack{i,j=1\\(i\ne j)}}^n \Delta y_i \Delta y_j\,,\non \\
 &=& \biggl(1 + \frac{1}{n}\biggr) \sum_{i=1}^n\Delta y_i^2 -  \frac{1}{n} \biggl(\sum_{i=1}^n\Delta y_i \biggr)^2\,,
\ea  
and the metric is
\be\n{2.10}
g_{ij} = 
\begin{pmatrix}
  1     & -1/n       & \cdots   &  -1/n \\
  -1/n  & 1          & \cdots   &  -1/n \\
\vdots & \vdots & \ddots & \vdots \\
  -1/n  &  -1/n    & \cdots     &  1 
\end{pmatrix}.
\ee
Using recursion and row reduction, or applying Sylvester's theorem, it is easy to see that the determinant is
\be\n{2.11}
g = n^{-n}(n+1)^{n-1}\,.
\ee
From \eq{2.2}, one obtains
\be\n{2.11a}
V_{\ind{FP}} = n^{-n/2}(n + 1)^{(n-1)/2} \ell^n
\ee
for the $n$-volume of the FP.

We now compute the covering radius $R$, which is the distance from the origin to the most distant point of the
WS centered at the origin. To find the boundary of the WS centered at the origin, we first find the equation of the plane that lies halfway between the origin and a lattice point with coordinates $(0,\cdots, 0,
\ell, \cdots, \ell)$, where the number of zeros is $k$ and the number
of $\ell$'s is $n-k$. We take this form for an FP vertex because it is sufficiently general, i.e.
according to the distance function form \eq{2.9}, the coordinates can be permuted without changing
the distance value. In contrast to the $\mathbb{Z}^n$ lattice, every FP vertex defines a WS boundary planes.  After multiplying the squared distance by an
overall factor of $n/(n+1)$, the coordinates in the planes satisfy the
equation
\ba
\non 
\sum_{i=1}^n y_i^2 & - &\frac{1}{n+1} \biggl( \sum_i y_i \biggr)^2 =
\sum_{i=1}^k y_i^2 + \sum_{i=k+1}^n (y_i-\ell)^2\\
&-& \frac{1}{n+1} \biggl( \sum_{i=1}^k y_i + \sum_{i=k+1}^n (y_i-\ell) \biggr)^2\,.
\ea
This expression can be simplified, rearranged, and divided by $2 \ell(n-k)/(n+1)$ to obtain
\be\non
\sum_{i=1}^n y_i = \frac{n+1}{n-k} \sum_{i=k+1}^n y_i - \frac{k+1}{2} \ell\,.
\ee
Writing the (unity) coefficient of the l.h.s. as $(n+1)/(n-k) - (k+1)/(n-k)$, and
canceling the common terms in the sums, gives
\be\non
\frac{n+1}{n-k} \sum_{i=1}^k y_i - \frac{k+1}{n-k} \sum_{i=1}^n y_i   = - \frac{k+1}{2} \ell\,.
\ee
Multiplying this expression by $(n-k)/(k+1)$ yields the following formula, which defines the planes bounding the WS cell:
\be\n{e:planeform}
\sum_{i=1}^n y_i =  \frac{n+1}{k+1} \sum_{i=1}^k y_i  +  \frac{n-k}{2} \ell\,.
\ee
Although we obtained this equation for a specific subset of
vertices, it is trivial to obtain the corresponding equation for any
vertex, by replacing the sum from $1$ to $k$ with a sum over any $k$
of the coordinates. Changing the sign of $\ell$ gives the
corresponding parallel plane bounding the WS on the other side of the
origin.  For this reason, the WS is sometimes called a ``permutohedron''
\cite{Conway} and denoted $P_n$.

To obtain the covering radius $R$, we intersect a set of $n$ bounding
planes defined by \eq{e:planeform}, to identify a point at this
radius in the WS.  The $k=0$ equation implies 
\be\n{2.14}
\sum_{i=1}^n y_i = \frac{n \ell}{2}\,.
\ee
The $k=1$ equation then implies $y_1=\ell/(n+1)$.  Combining these
with the $k=2$ equation implies $y_2=2\ell/(n+1)$. Continuing in this
fashion, intersecting all of the planes implies $y_i = i \ell/(n+1)$.
The squared covering radius of the WS is thus given by
\ba\n{2.15}
R^2 & = &\left(1+\frac{1}{n}\right)\sum_{i=1}^n y_i^2  - \frac{1}{n} \biggl( \sum_{i=1}^n y_i \biggr)^2\non\\
& = &\left(1 + \frac{1}{n}\right)\frac{\ell^2}{(n+1)^2} (1^2 + \cdots + n^2) - \frac{1}{n}\left(\frac{n \ell}{2}\right)^2\non\\
& = &  \frac{\ell^2}{n(n+1)} \frac{n(n+1)(2n+1)}{6} - \frac{n \ell^2}{4}\non\\
& = & \frac{1}{12}(n+2) \ell^2\,.
\ea
As before, we can express the WS $n$-volume in terms of $R$:
\be
\label{e:ws-volume}
V^{A_n^*}_{\ind{WS}} = \left[ \frac{12(n+1)}{n(n+2)} \right]^{n/2} (n+1)^{-1/2} R^n.
\ee
This will be useful later, when we compare lattices at fixed WS volume.

\section{\label{s:lossfraction} The fraction of lost detections}

A template bank is discrete, so most points in parameter space do not
have an exactly matching template.  As a result, there is the
detection mismatch, which, on average, results in lost detections.
Here, $N_{D}$ denotes the total number of sources detectable above a
certain SNR threshold, and $N_{\ind{lost}}$ is the number of lost
detections, in comparison with a closely spaced (ideal) bank that catches all
signals.

The fraction of lost detections depends upon the effective
dimensionality $d$ of the source distribution.  If sources are
uniformly placed in a 3-dimensional Euclidean space, then the number
of sources $N$ grows as the distance $L$ as $dN \propto L^2 dL$. Similarly,
if they are arranged in a 2-dimensional plane (for example, a thin Galactic disk) 
then $dN \propto L dL$.  So here we define $d$ by $dN \propto
L^{d-1} dL$ and assume that the squared SNR is proportional to
$1/L^2$.

If the volume of parameter space is much larger than a WS cell and the
template bank is a lattice, then 
the fraction of lost
detections in the spherical approximation \cite{AllenSpherical} is given by \cite{NewAllenPaper}
\be
\frac{N_{\ind{lost}}}{N_{D}} \approx  \frac{1}{V_{\ind{WS}}}\int_{\ind{WS}} f(r) \, dV,
\ee
where the integral is over a single WS cell, and 
the integrand is
\be
\label{e:fullmismatch}
f(r) =
\begin{cases}
 1 - \cos^d r  & \text{for } r \le \pi/2 \\
 1             & \text{for } r > \pi/2 
\end{cases}.
\ee
The ratio $N_{\ind{lost}}/N_{D}$ defines the ``loss fraction'' of the
lattice, i.e. the fraction of potentially-detectable signals
which the lattice fails to catch.  Equivalently,
$1-N_{\ind{lost}}/N_{D}$ is the efficiency of the lattice: the
expected fraction of potentially detectable signals which are indeed
found.

Provided that the WS cell is not too large, so that $R<\pi/2$, the integrand can be expanded in a series, giving a loss fraction
\ba\n{3.1}
\frac{N_{\ind{lost}}}{N_{D}}&\approx&\frac{1}{V_{\ind{WS}}}\int_{\ind{WS}}\left(1-\cos^d(r)\right)\, dV \non\\
&=&\frac{d}{2}\langle r^2\rangle-\frac{d(3d-2)}{24}\langle r^4\rangle+\frac{d(15d^2-30d+16)}{720}\langle r^6\rangle\non\\
&-&\frac{d(105d^3-420d^2+588d-272)}{40320}\langle r^8\rangle+\cdots\,.
\ea
Here,
\be\n{3.2}
\langle r^p\rangle = \frac{1}{V_{\ind{WS}}}\int_{\ind{WS}}r^p \, dV
\ee
denotes the normalized $p$'th moment of the lattice.

Provided that the effective dimensionality of the source distribution
$d> 8/\pi^2 \approx 0.81$, the quadratic approximation always implies
a larger fraction of signals lost than the spherical approximation,
because $1-\cos^d(r) < r^2\, d/2$ on the interval $r\in [0, \pi/2]$.

Appendix \ref{s:zlatticemoments} shows how the even moments may be computed for the
$\mathbb{Z}^n$ lattice. 
 The first six of these, which suffice for this paper, are
\ba\n{3.3} \langle
r^2\rangle&=&\frac{n\ell^2}{12}\,, \non \\
\langle r^4\rangle& =& \frac{n\ell^4}{720}(5n+4)\,,\non\\
\langle r^6\rangle&=&\frac{n\ell^6}{60480}(35n^2+84n+16)\,,\\
\langle r^8\rangle&=&\frac{n\ell^8}{3628800}(175n^3+840n^2+656n-96)\,,\non\\
\langle r^{10}\rangle&=&\frac{n\ell^{10}}{95800320}(385n^4+3080n^3+5456n^2\non\\
&+&352n-768)\,,\non \\
\langle r^{12}\rangle&=&\frac{n\ell^{12}}{523069747200}(175175n^5+2102100n^4\non\\
\non
&+&6646640n^3+3747744n^2-2883712n + 35328)\,.
\ea We note that all of these quantities can be re-expressed in terms
of the covering radius $R^2=n \ell^2/4$.  The corresponding even
moments for the $A_n^*$ lattice are computed in Appendix
\ref{s:AnStarMoments}, but not repeated here.

In the following we shall consider $d=2$ and $d=3$ dimensional source distributions.

\subsection{The $\mathbb{Z}^n$ lattice: $d=2$ case}

\begin{table}
  \caption{\label{tab1} The maximal fraction of lost detections
    $N_{\ind{lost}}/N_{D}$ for the $\mathbb{Z}^n$ and $A_n^*$
    lattices in small dimensions $n$, for $d=2$ and $d=3$ dimensional
    source distributions. Note that all lattices have ``maximal'' WS
    radius $R=\pi/2$, which means that at a fixed dimension $n$, the
    WS cells have smaller volume for $\mathbb{Z}^n$ than for $A_n^*$.}
\begin{ruledtabular}
  \begin{tabular}{ccccc}
     & $\mathbb{Z}^n$ & $\mathbb{Z}^n$ & $A_n^*$ & $A_n^*$ \\
 n   & $d = 2$        & $d = 3$        & $d = 2$ & $d = 3$ \\
    \hline 
 2   & 0.558          & 0.665          & 0.642   & 0.736  \\           
 3   & 0.579          &  0.697         & 0.720   & 0.816  \\
 4   & 0.589          &  0.714         & 0.771   & 0.863  \\
 5   & 0.595          &  0.724         & 0.806   & 0.893  \\
 6   & 0.599          &  0.731         & 0.832   & 0.914  \\
 7   & 0.602          &  0.736         & 0.851   & 0.928  \\
 8   & 0.605          &  0.740         & 0.867   & 0.939  \\
 9   & 0.606          &  0.743         & 0.880   & 0.948  \\
 10  & 0.608          &  0.745         & 0.890   & 0.954  \\
 11  & 0.609          &  0.747         & 0.899   & 0.960  \\
 12  & 0.610          &  0.749         & 0.906   & 0.964  \\
 $n\to \infty$
     & 0.620          &  0.766         & 1       & 1
 \end{tabular}
\end{ruledtabular}
\end{table} 

\begin{figure*}[htb]
\begin{center}
\ba
&&\hspace{0cm}\includegraphics[width=6cm]{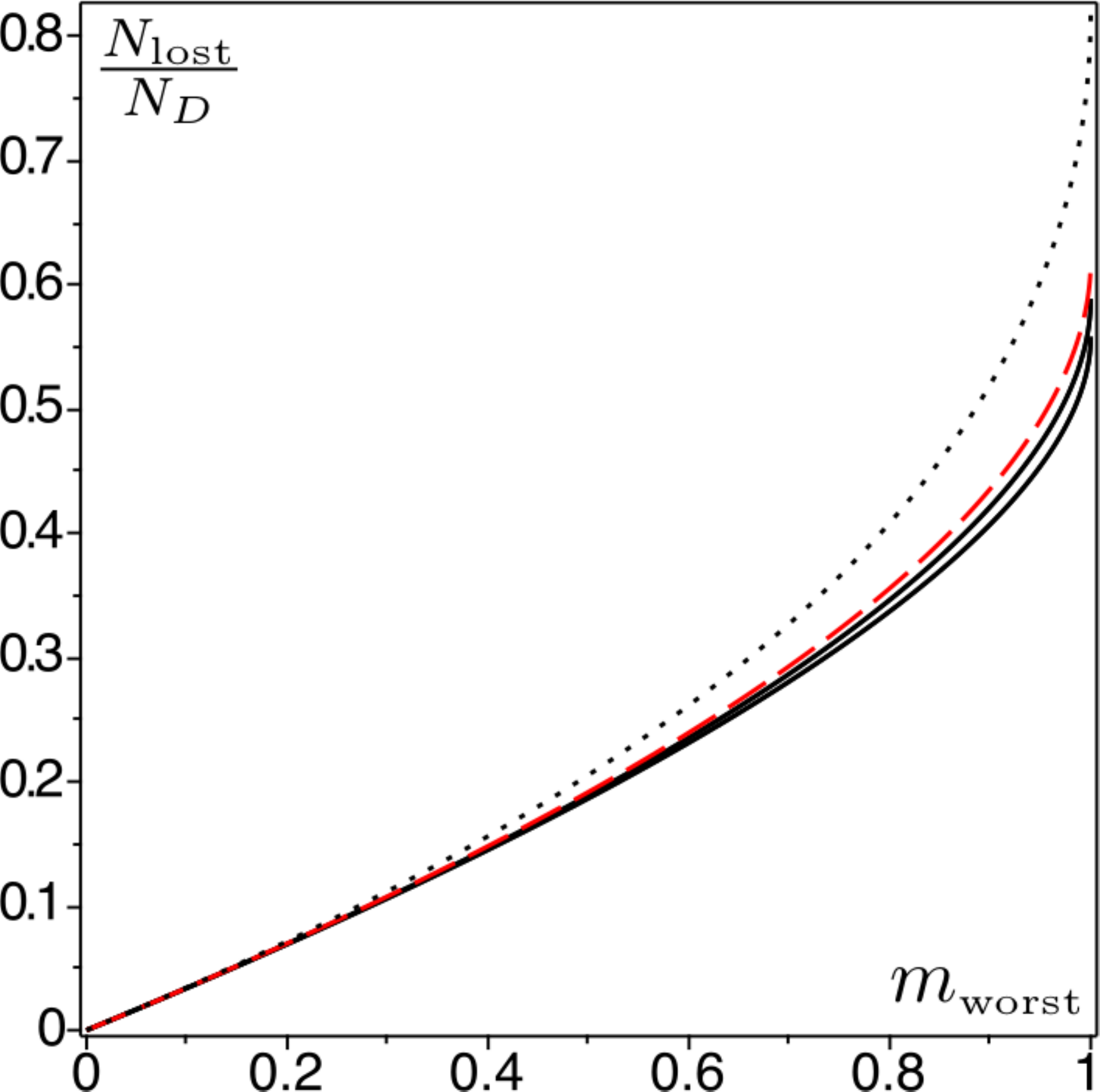}
\hspace{3cm}\includegraphics[width=6cm]{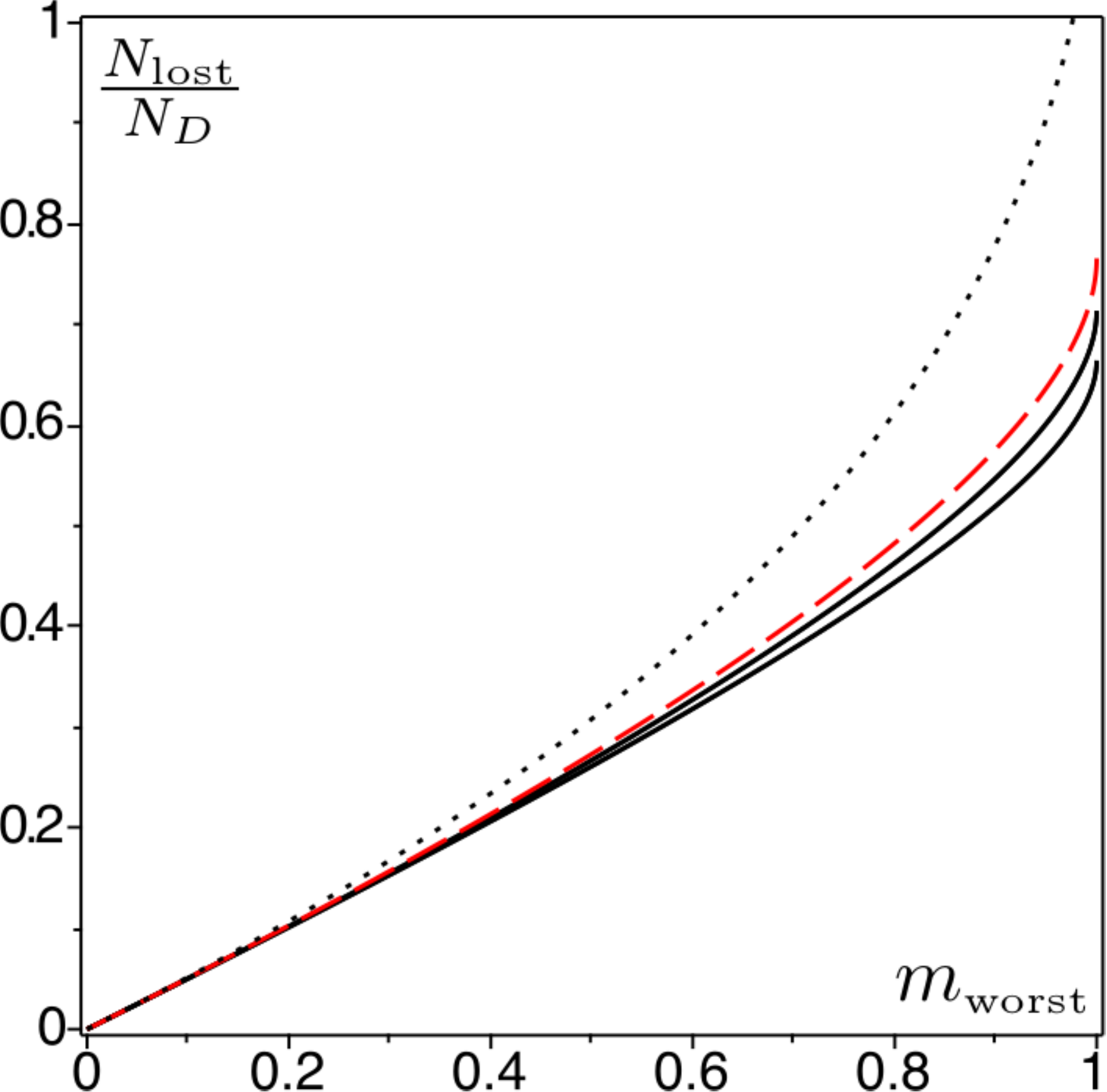}\non\\
&&\hspace{3cm}({\bf a})\hspace{8.5cm}({\bf b})\non
\ea
\caption{The fraction of lost detections for the $\mathbb{Z}^n$
  lattice in dimensions $n=2$ (lower solid curve) and $n=4$ (upper
  solid curve). For larger $n$ the corresponding curves group very
  close together; the red dashed curve shows the $n \to \infty$ limit
  of \eq{3.8}.  The left-hand plot shows a $d=2$~dimensional
  source distribution and the right-hand plot shows a
  $d=3$~dimensional distribution. The fraction of lost detections
  depends upon the spacing of the template bank, which is set by the
  covering radius $R$; in the spherical approximation \cite{AllenSpherical}, the worst
  mismatch $m_{\ind{worst}}=\sin^2R$.  For closely spaced templates
  (small $R$) no detections are lost.  For comparison the quadratic
  approximation \eq{3.5} is shown as a dotted curve. It predicts
  more lost signals than the spherical approximation suggests. The
  fraction of lost detections at maximum mismatch $R=\pi/2$ is given
  in Table~\ref{tab1}. As $n \to \infty$ and at covering radius
  $R=\pi/2$, about 62\% of detections are lost for a $d=2$~dimensional
  source distribution, and about 77\% are lost for a $d=3$~dimensional
  source distribution.
  \label{f2} }
\end{center}
\end{figure*}

\begin{figure*}[htb]
\begin{center}
\ba
&&\hspace{0cm}\includegraphics[width=6cm]{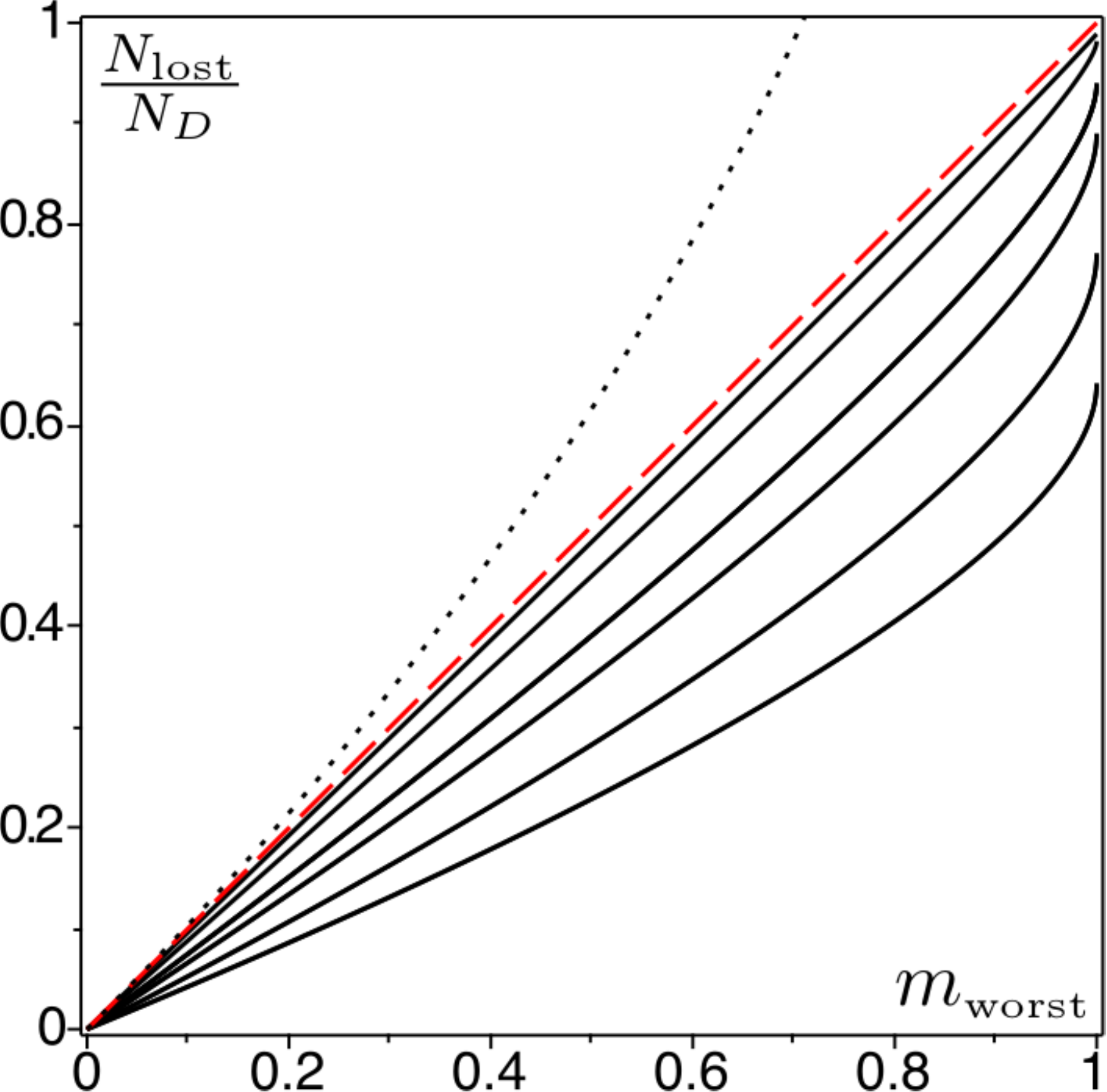}
\hspace{3cm}\includegraphics[width=6cm]{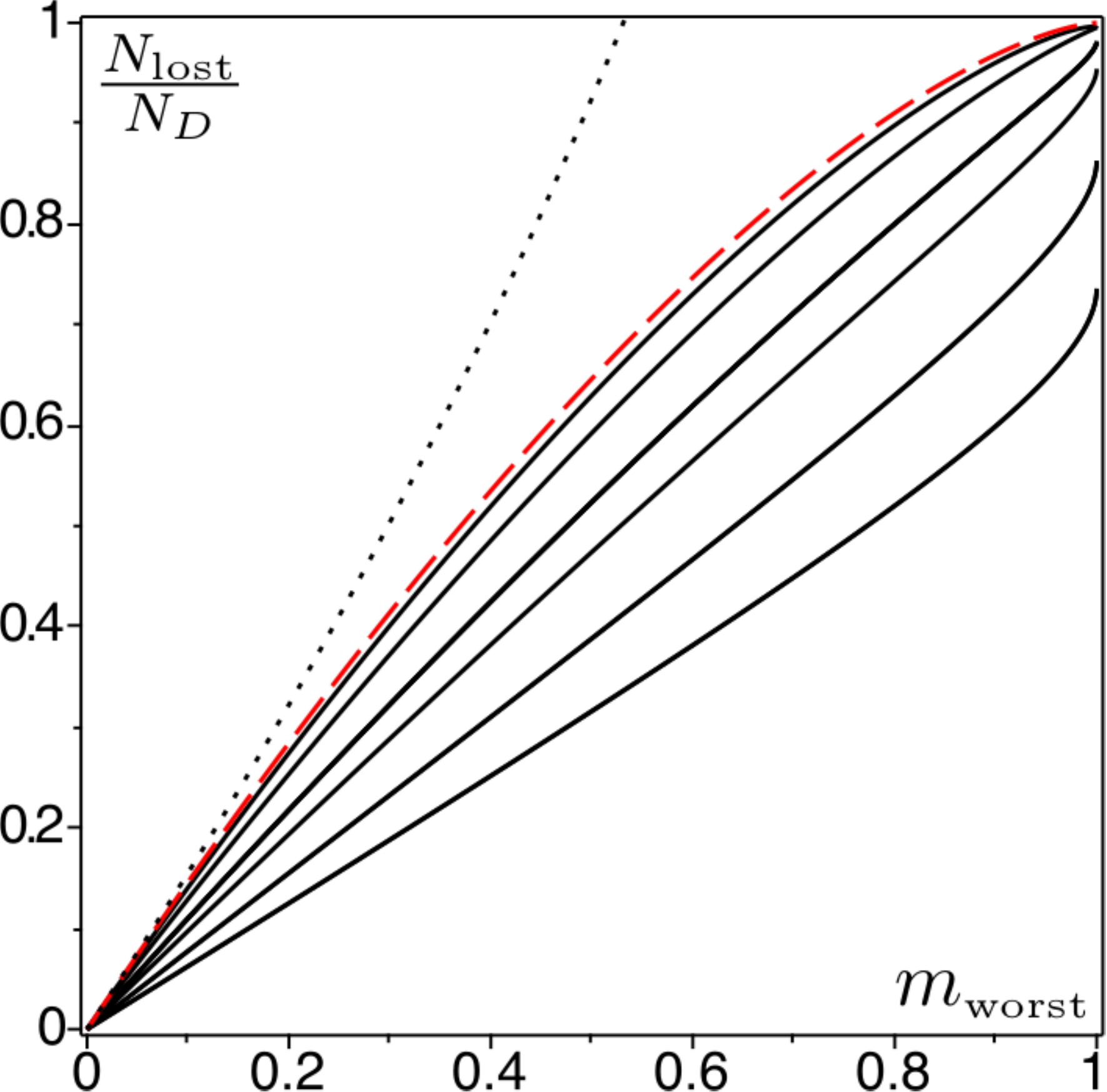}\non\\
&&\hspace{3cm}({\bf a})\hspace{8.5cm}({\bf b})\non
\ea
\caption{The fraction of lost detections for the $A_n^*$ lattice in
  different numbers of dimensions (moving upwards)
  $n=2,4,10,20,100,1000$.  The left-hand plot shows a
  $d=2$~dimensional source distribution and the right-hand plots shows
  a $d=3$~dimensional distribution.  As in the previous figure, the $n
  \to \infty$ limit is shown in red, and the plot is restricted to
  $R\leq\pi/2$.  The quadratic approximation is shown as a dotted
  curve in the limit $n\to\infty$ [see \eq{3.9}]. The fraction of lost
  detections at maximum mismatch $R=\pi/2$ is given in
  Table~\ref{tab1}.
 \label{f3}}
\end{center}
\end{figure*}

For a source distribution with effective dimensionality $d=2$, we now
evaluate the fraction of lost sources, assuming that the covering radius $R
\le \pi/2$.  The integrand of \eq{3.1} ($f(r)=\sin^2r$, the
mismatch in the spherical approximation \cite{AllenSpherical}) is approximated (within
1\% ) by taking terms up to the eighth moment.  Then, \eq{3.1} takes the form
\ba\n{3.4}
\non
\frac{N_{\ind{lost}}}{N_{D}}& \approx & \langle r^2\rangle-\frac{1}{3}\langle r^4\rangle+\frac{2}{45}\langle r^6\rangle - \frac{1}{315}\langle r^8\rangle \\
\non
& = & \frac{1}{3} R^2 - \frac{5n + 4}{135n} R^4 + \frac{70n^2 + 168n +32}{42525n^2}  R^6 \\
&& - \frac{175n^3+840n^2+656n-96}{4465125n^3} R^8.
\ea
We plot this quantity in
Fig.~\ref{f2}(a), where $m_{\ind{worst}} =\sin^2 R$ denotes the worst-case mismatch in the spherical approximation.

Fig.~\ref{f2}(a) also compares the spherical approximation \cite{AllenSpherical} to the mismatch
with the prediction one would find using the normal quadratic
approximation.  If the lattice is widely spaced (sparse), then the
spherical approximation predicts significantly fewer lost signals than
the standard quadratic approximation.  The quadratic approximation
keeps only the first term in \eq{3.4}, so
\be\n{3.5}
\left[ \frac{N_{\ind{lost}}}{N_{D}}\right]_{\text{Quadratic-Approximation}}=\frac{1}{3}\arcsin^2(\sqrt{m_{\ind{worst}}})\, ,
\ee
which is valid in any dimension $n$.  To enable a fair comparison with
the spherical approximation, we need to examine the two expressions
for the same lattice, meaning at the same WS radius $R$.  So in
\eq{3.5}, this is still related to the worst-case mismatch via
the spherical approximation \cite{AllenSpherical}
$m_{\ind{worst}} =\sin^2 R$ (rather than with the
quadratic approximation $m_{\ind{worst}} = R^2$).  

Results of numerical computations of the maximal fraction of lost
detections are presented in Table~\ref{tab1}.

\subsection{The $\mathbb{Z}^n$ lattice: $d=3$ case}

For $d=3$ the integrand in \eq{3.1} is $f(r)=1-\cos^3r$, and we again assume $R \le \pi/2$. The expression \eq{3.1} takes the following form:
\ba\n{3.7}
\frac{N_{\ind{lost}}}{N_{D}}&\approx&\frac{3}{2}\langle r^2\rangle-\frac{7}{8}\langle r^4\rangle+\frac{61}{240}\langle r^6\rangle-\frac{547}{13440}\langle r^8\rangle\non\\
&+&\frac{703}{172800}\langle r^{10}\rangle-\frac{44287}{159667200}\langle r^{12}\rangle\,.
\ea
Here, to maintain 1\% accuracy in the integrand we have had to include
more terms than for $d=2$.  Fig.~\ref{f2}(b) illustrates how the
fraction of lost detections depends on the covering radius (via the
worst-case mismatch $m_{\ind{worst}}$). In the case of quadratic approximation we keep only
the first term in \eq{3.7},
\be\n{3.7a}
\left[ \frac{N_{\ind{lost}}}{N_{D}}\right]_{\text{Quadratic-Approximation}}=\frac{1}{2}\arcsin^2(\sqrt{m_{\ind{worst}}})\,,
\ee
valid in any dimension $n$. For a widely spaced lattice the spherical approximation predicts significantly fewer lost signals than the standard quadratic approximation. The worst-case values
(fraction of lost detections at WS radius $R=\pi/2$) are shown in
Table~\ref{tab1}.

\subsection{The $A_n^*$ lattice: $d=2$ and $d=3$ cases}

As for the $\mathbb{Z}^n$ lattice, we can again estimate how the
fraction of lost detections depends upon the covering radius. For the
$A_n^*$ lattice, we can compute the moments $\langle r^p\rangle$
exactly, but cannot give a closed analytic form as we did for
the $\mathbb{Z}^n$ lattice.  We use the exact expressions obtained in
Appendix~\ref{s:AnStarMoments}, and substitute these into the
expressions \eq{3.4} and \eq{3.7}. The plots of the fraction of lost
detections versus $m_{\ind{worst}}$ are given in Fig.~\ref{f3}, and
some worst-case values are shown in Table~\ref{tab1}.

\section{\label{s:largen} Large $n$ limits}

The reader will notice that as the dimension $n$ of the parameter
space gets large, the curves appear to approach a limit.  This is
explained in Sec.~\ref{s:mismatch}, where we show that as $n$ gets large,
the mismatch distribution function becomes sharply peaked at $r^2 =
R^2/3$ for the $\mathbb{Z}^n$ lattice and at $r^2=R^2$ for the $A_n^*$ lattice. 
Thus, for the $\mathbb{Z}^n$ lattice, \eq{3.1} immediately gives
\ba\n{3.8}
\non
\lim_{n \to \infty} \frac{N_{\ind{lost}}}{N_{D}} & = & 1-\cos^d\left({R}/{\sqrt{3}}\right)\,\\
& = & 1 - \cos^d\left(\frac{\arcsin \sqrt{m_{\ind{worst}}}}{\sqrt{3}}  \right),
\ea
where we have used the relationship $m_{\ind{worst}} = \sin^2 R$
between the WS radius and the worst-case mismatch.

For a source distribution with effective dimensionality $d=2$ this has
a limiting value of $N_{\ind{lost}}/N_{D} \approx 0.620$ for
$m_{\ind{worst}} =1$. So, if there are at least a few dimensions to
parameter space, then placing templates in a rectangular grid at unit
mismatch will recover about 38\% of signals.  For a source
distribution with effective dimensionality $d=3$, the limiting value
is $N_{\ind{lost}}/N_{D} \approx 0.766$, so a rectangular grid
at unit mismatch would recover about 23\% of signals.

In the case of the $A_n^*$ lattice, Sec.~\ref{s:mismatch} shows that in the limit of large $n$ we have
\be
\lim_{n\to\infty}\langle r^2\rangle_{\ind{WS}} = R^2\,,
\ee
where the covering radius $R\in(0,\pi/2]$.
In fact this is also true for the higher moments, as can be seen
from either Sec.~\ref{s:mismatch}  or from the results of Appendix~\ref{s:AnStarMoments}, meaning that 
\be
\lim_{n\to\infty}\langle r^{2m}\rangle_{\ind{WS}} = R^{2m}\hhh m=1,2,3,...\,\,.
\ee
Thus,
\be
\label{e:bignforAstar}
\lim_{n \to \infty} \frac{N_{\ind{lost}}}{N_{D}} =1-\cos^d R \,,
\ee
which leads to a worst-case limit of unity, as shown in Table~\ref{tab1}.
For large dimensions the quadratic approximation of the fraction of lost detections can also be constructed in the closed form [cf. \eq{3.5}],
\be\n{3.9}
\lim_{n \to \infty}\left[\frac{N_{\ind{lost}}}{N_{D}}\right]_{\text{Quadratic-Approximation}}=\frac{d}{2}\arcsin^2(\sqrt{m_{\ind{worst}}})\,.
\ee
This expression is shown by the dotted curves in Fig.~\ref{f3}.

\begin{figure*}[htb]
\begin{center}
\ba
&&\hspace{0cm}\includegraphics[width=8cm]{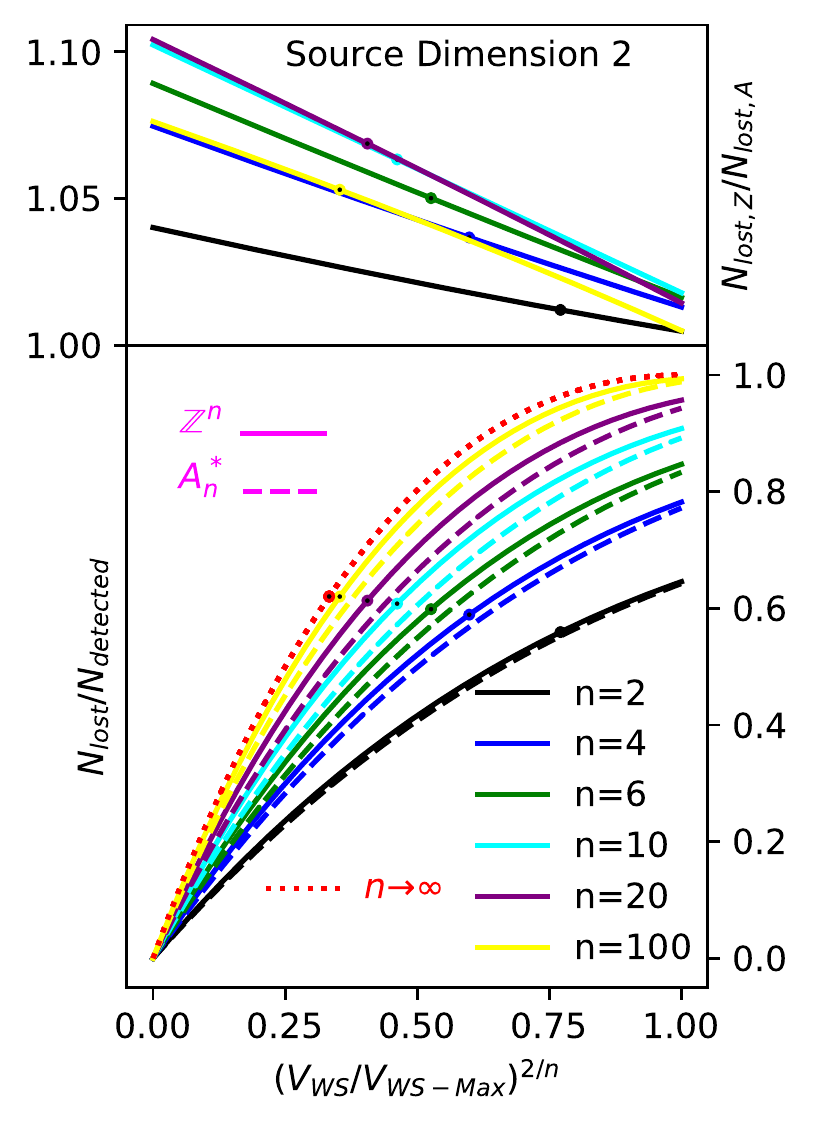}
\hspace{1cm}\includegraphics[width=8cm]{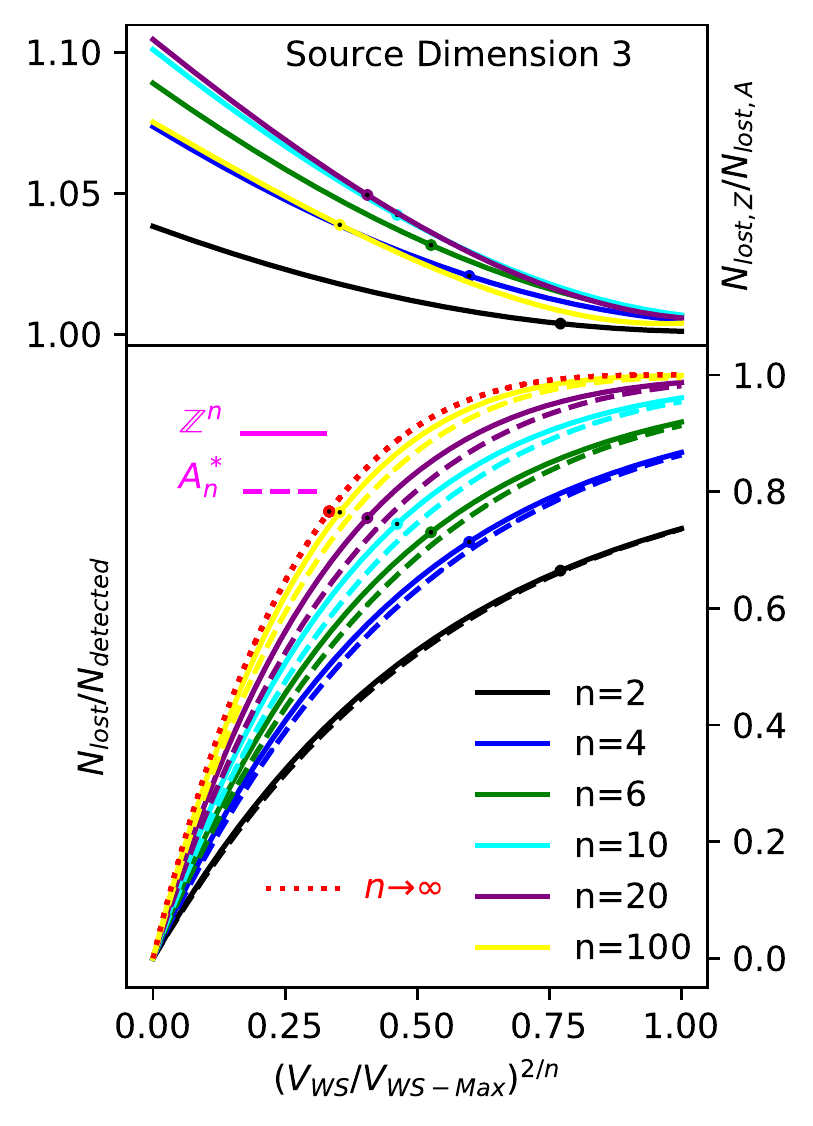}\non\\
&&\hspace{4cm}({\bf a})\hspace{8.5cm}({\bf b})\non
\ea
\caption{\label{f4} A comparison of lattice loss fractions at fixed
  computing cost.  Lower curves: the loss fractions
  $N_{\ind{lost}}/N_{D}$ for the $\mathbb{Z}^n$ and $A_n^*$ lattices,
  at fixed WS cell volume $V_{\ind{WS}}$. The red dotted curves represent 
  the lattice loss fraction in the limit $n\to\infty$ [see \eq{fr:limit}]. 
  Upper curves: ratios of
  these loss fractions.  The horizontal axis normalization is given by
  \eq{e:axisdef}.  The dots on the curves indicate where the
  covering radius for $\mathbb{Z}^n$ reaches $R=\pi/2$. The left
  (right) plots are for a $d=2$ ($d=3$)~dimensional source
  distributions.}
\end{center}
\end{figure*}

\section{\label{s:compare} Comparison of  $\mathbb{Z}^n$ and $A_n^*$ at fixed computing cost}

To evaluate the relative loss fractions of the $\mathbb{Z}^n$ and $A_n^*$
lattices at fixed computing cost, we must compare them for identical
values of the WS cell volume $V_{\ind{WS}}$. This ensures that the same
number of templates would be employed to cover a given volume of
parameter space.

Such a comparison is
shown in Fig.~\ref{f4}.  The horizontal axis ``$x$'' in these plots is
proportional to the ``squared length'' 
$V_{\ind{WS}}^{2/n}$.  In the figure, this is normalized to reach unity when the
covering radius of the $A_n^*$ lattice reaches $R=\pi/2$.
From \eq{e:ws-volume}, the resulting normalization factor is the inverse of
\ba
\label{e:axisdef}
\non
V_{\ind{WS-Max}}^{2/n} & = & \biggl[ V_{\ind{WS}}^{A_n^*}(R=\pi/2) \biggr]^{2/n}\\ 
& = & \frac{3 \pi^2 (n+1)}{n(n+2)} (n+1)^{-1/n}.
\ea
Thus, if we denote the horizontal axes of Fig.~\ref{f4} by $x=(V_{\ind{WS}}/V_{\ind{WS-Max}})^{2/n}$, 
by using \eq{volZ} and \eq{e:ws-volume} we have
\be\n{e:xdef}
x =\left\{
\begin{array}{ll}
\cfrac{4}{3 \pi^2}\left( \cfrac{n+2}{n+1} \right) (1+n)^{1/n} R^2  & \mbox{for $\mathbb{Z}^n$} \\
&\\
\cfrac{4R^2}{\pi^2} & \mbox{for $A_n^*$}\,,
\end{array}
\right.
\ee
where $R$ is WS cell covering radius of the corresponding lattice. Note
that when the two lattices are compared at a given point on the
$x$-axis, they have equal WS cell volume, hence they have {\it
  different} WS radii, and correspondingly {\it different} values of
$\ell$.

At fixed $V_{\ind{WS}}$, the WS radius $R$ of the $\mathbb{Z}^n$
lattice is always larger than the WS radius of the $A_n^*$ lattice.
Since we allow the WS radius for $A_n^*$ to reach maximal value
$\pi/2$, it follows that in the plots in Fig.~\ref{f4}, the WS radius of
$\mathbb{Z}^n$ exceeds $\pi/2$ for some of the domain.  The transition point
where the WS radius of $\mathbb{Z}^n$ reaches $R=\pi/2$  
is denoted by a dot on
the curves; to the right of this dot, the mismatch of the
$\mathbb{Z}^n$ lattice is set to unity for $r>\pi/2$ in accordance with \eq{e:fullmismatch}. Thus, to the
right of this dot, the $\mathbb{Z}^n$ results have been obtained with
Monte Carlo integration, since the analytic formulae obtained earlier
only hold for $R\le \pi/2$.  As can be seen from \eq{e:xdef}, the location of this dot approaches
$x = 1/3$ in the large-$n$ limit.

One can see that these plots have taken us away from the quadratic
approximation to the mismatch. To get some sense of how far away,
consider the maximum mismatch at the locations of the dots.  In the
quadratic approximation, this would be $m=r^2 = \pi^2/4 \approx 2.47$,
more than double the maximum allowed value of $m=1$.  In the quadratic
approximation to the mismatch, the lower curves of Fig.~\ref{f4} would
be straight lines tangent to the given curves at $V_{\ind{WS}}=0$.
The upper curves would be horizontal lines passing through the
$V_{\ind{WS}}=0$ values.

The results of \cite{NewAllenPaper} show that for small mismatch,
where the quadratic approximation applies, the $A_n^*$ lattice is
only slightly less lossy than the $\mathbb{Z}^n$ lattice.  We can
now see that this marginal advantage {\it decreases} for larger mismatch:
the upper part of Fig.~\ref{f4} shows the ratio of the loss fractions
for the two lattices.  The efficiency of the $A_n^*$ lattice is at
most $\approx 10$~\%  higher than that of the $\mathbb{Z}^n$ lattice.

The large-$n$ limits of Sec.~\ref{s:largen} are informative and can be
easily evaluated.  Taking $n \to \infty$ in \eq{e:xdef} the loss
fractions \eq{3.8} and \eq{e:bignforAstar} for both the lattices
take the identical form
\be\n{fr:limit}
\frac{N_{\ind{lost}}}{N_{D}} = 1 - \cos^d \left( \frac{\pi}{2} \sqrt{x} \right).
\ee
This is shown by the dotted red curves in Fig.~\ref{f4}. The
transition point $x=1/3$ is indicated with a dot; at that point the
covering radius of the $\mathbb{Z}^n$ lattice equal to $\pi/2$.  While
the $A_n^*$ lattice has the same curve, the transition is only
relevant for the $\mathbb{Z}^n$ lattice.  For large $n$, the ratio of
the loss fractions approaches unity, as can be seen from
\eq{fr:limit}.

\section{\label{s:mismatch} Distribution function of the squared distance}

To understand and interpret the results presented above, it is helpful to define the
{\it mismatch distribution function} $P_m(m)$.  This is defined as a
probability distribution: if points in parameter space are chosen
``at random'' then the probability that the mismatch lies in the range
$(m, m+dm)$ is $P_m(m) dm$.  Here, we compute $P_m(m)$ under the
assumption that the probability of selecting a particular point in
parameter space is a uniform distribution in the lattice coordinates $y_i \in
[0,\ell]$.  This is equivalent to a uniform distribution in $x_i$.

In the quadratic and spherical approximations \cite{AllenSpherical}, the mismatch is a
one-to-one function of the squared distance $r^2$, assuming of course
in the spherical case that we restrict attention to $r\in[0,\pi/2]$.  Hence,
the mismatch distribution can be obtained from the radius distribution
function $P_{r^2}(r^2)$, assuming the same uniform distribution of the
$y _i$.  This distribution function can be used to compute an average value
of an integrable function $f$ of $r^2$, 
\be
\hspace{-0.2cm}\langle f(r^2) \rangle=\frac{1}{V_{\ind{WS}}}\int_{\ind{WS}}\hspace{-0.1cm}f(r^2)dV=\int_0^{R^2}\hspace{-0.2cm}f(r^2)P_{r^2}(r^2)dr^2\,.
\ee
Thus, the quantity we wish to compute is the distribution of
the values of the quadratic forms given in \eq{2.4} for the
$\mathbb{Z}^n$ lattice and in \eq{2.9} for the $A_n^*$ lattice.

\subsection{$r^2$-distribution for the $\mathbb{Z}^n$ lattice}

For finite values of the dimension $n$ we have not found a simple
closed form for $P_{r^2}(r^2)$, although we can give expressions for
$n=1,2$, and $3$.  However, the large-$n$ limit is easily computed.

\begin{figure*}
  \begin{center}
    \ba
    \non
    &&\hspace{-0.02\textwidth}\includegraphics[width=0.55\textwidth]{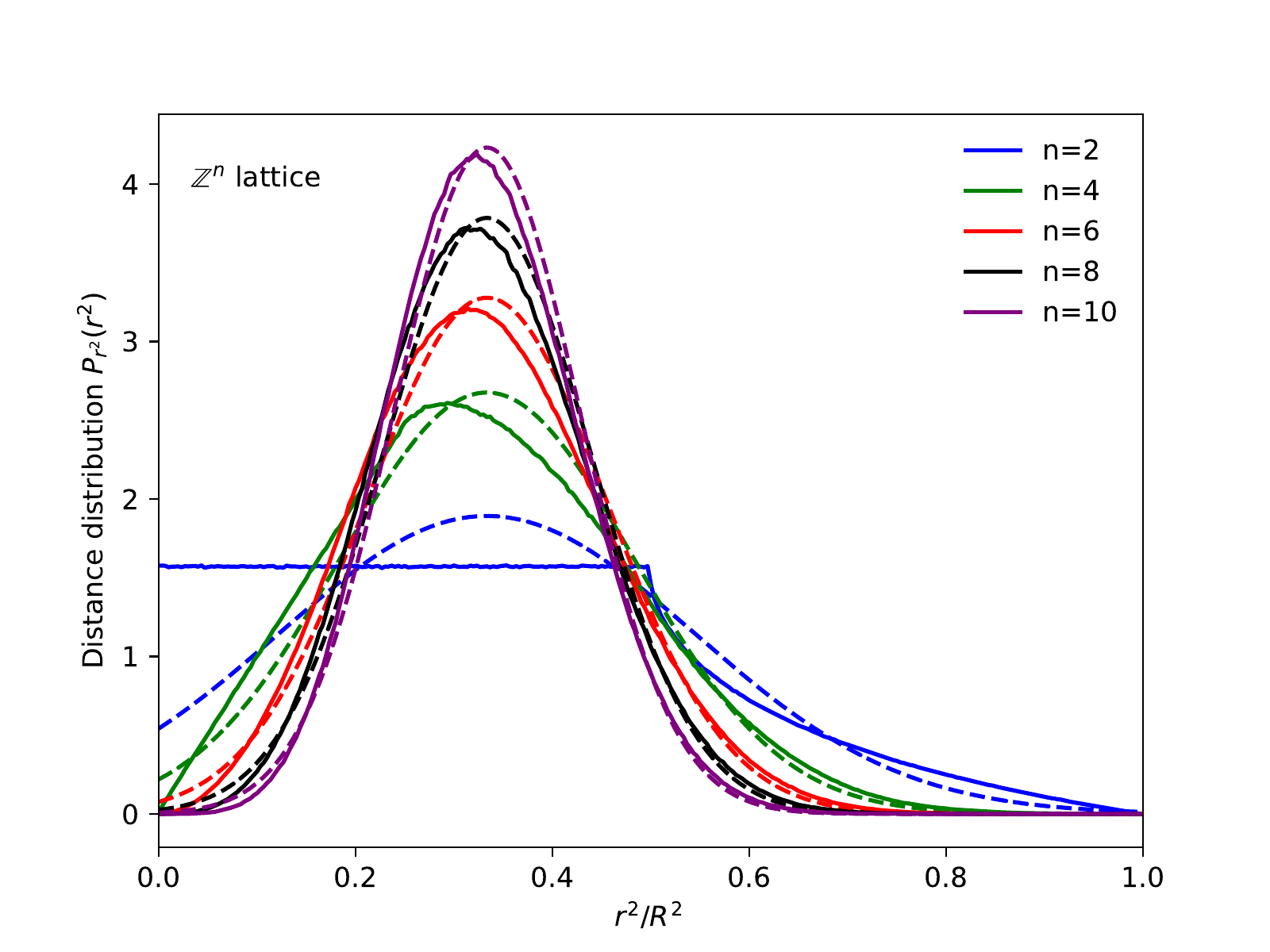}
    \hspace{-0.04\textwidth}\includegraphics[width=0.55\textwidth]{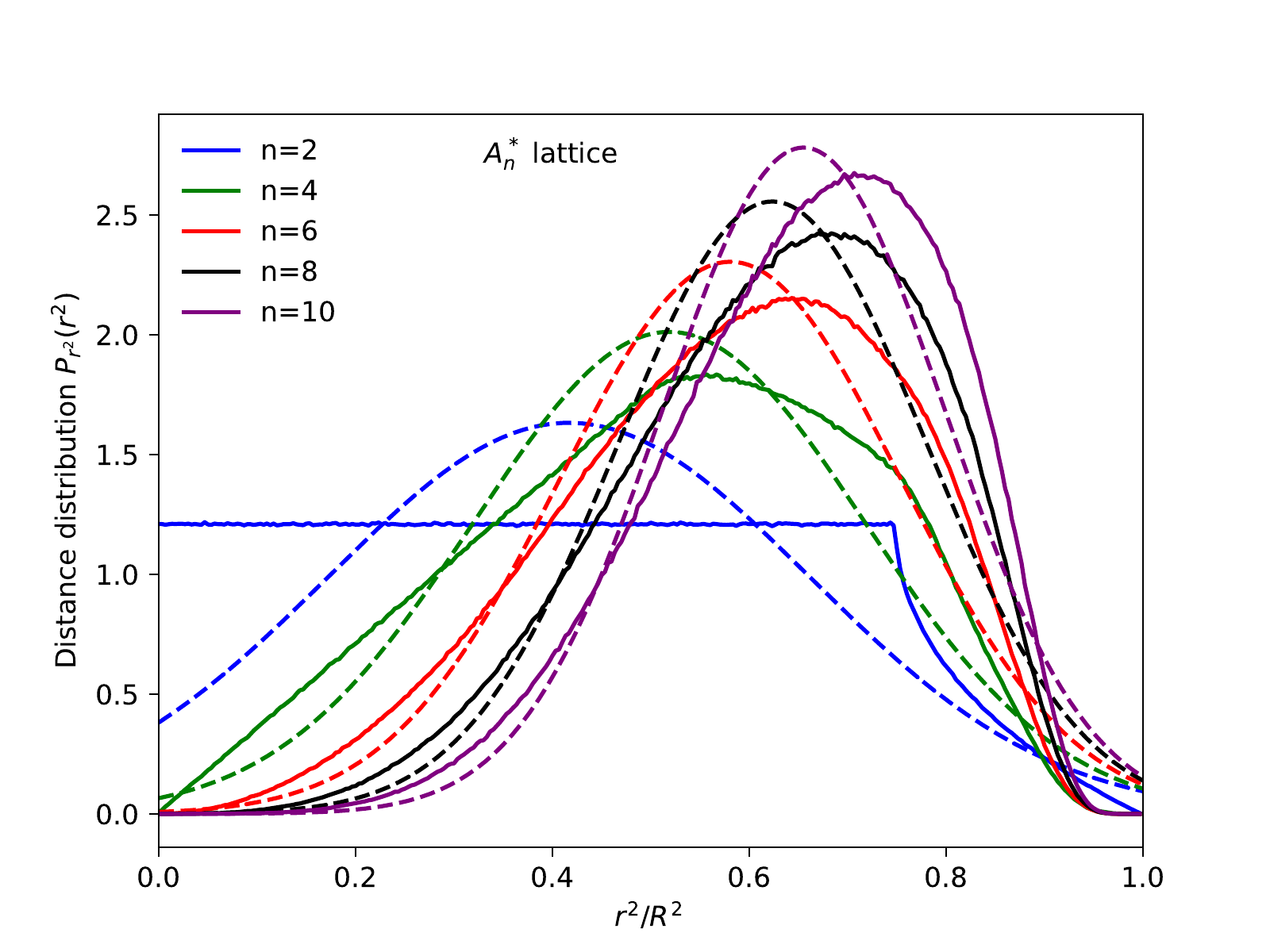}\\
    \non
    &&\vspace{0.0cm}\non\\
    &&\hspace{-0.02\textwidth}\includegraphics[width=0.55\textwidth]{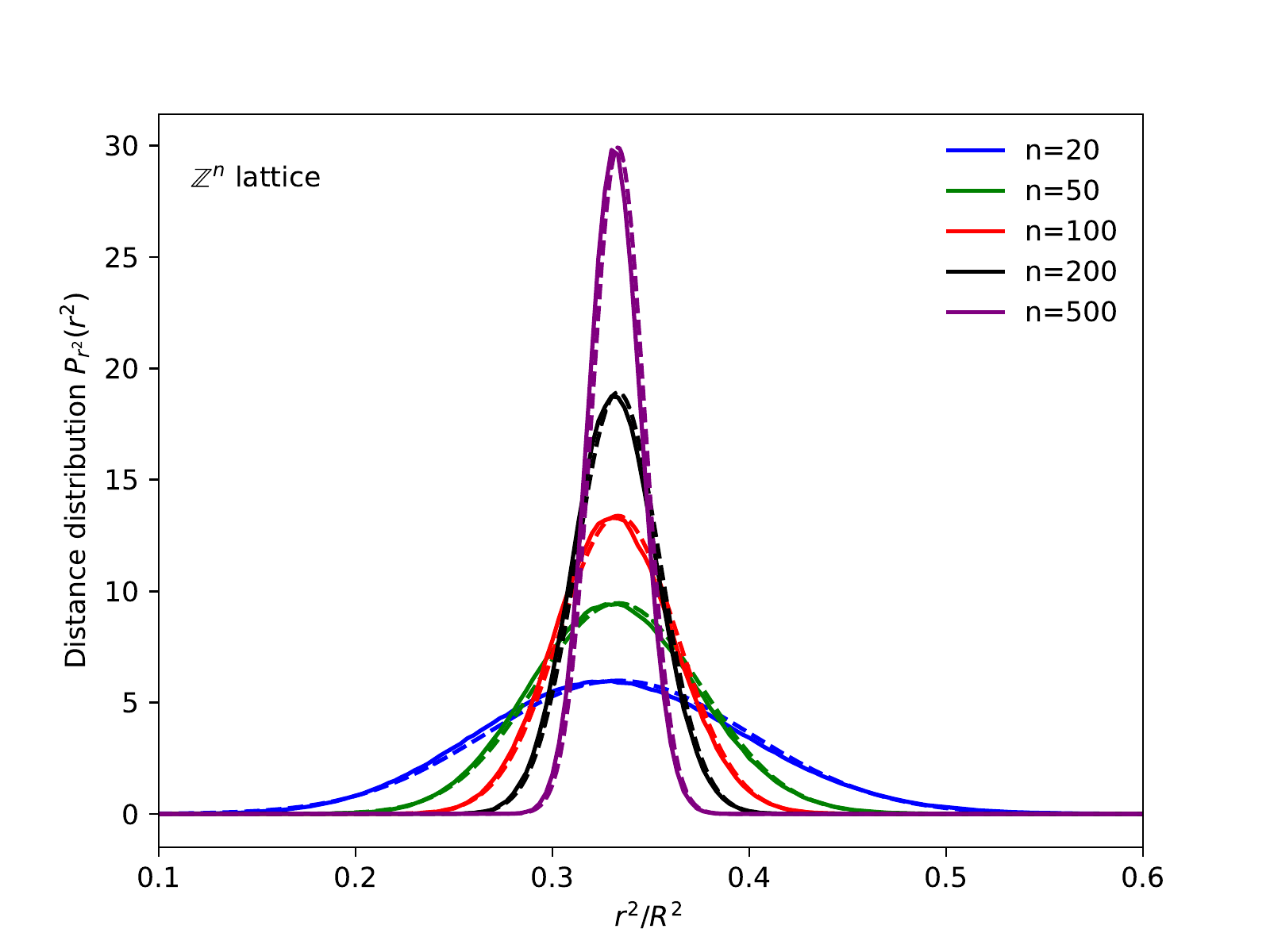}
    \hspace{-0.04\textwidth}\includegraphics[width=0.55\textwidth]{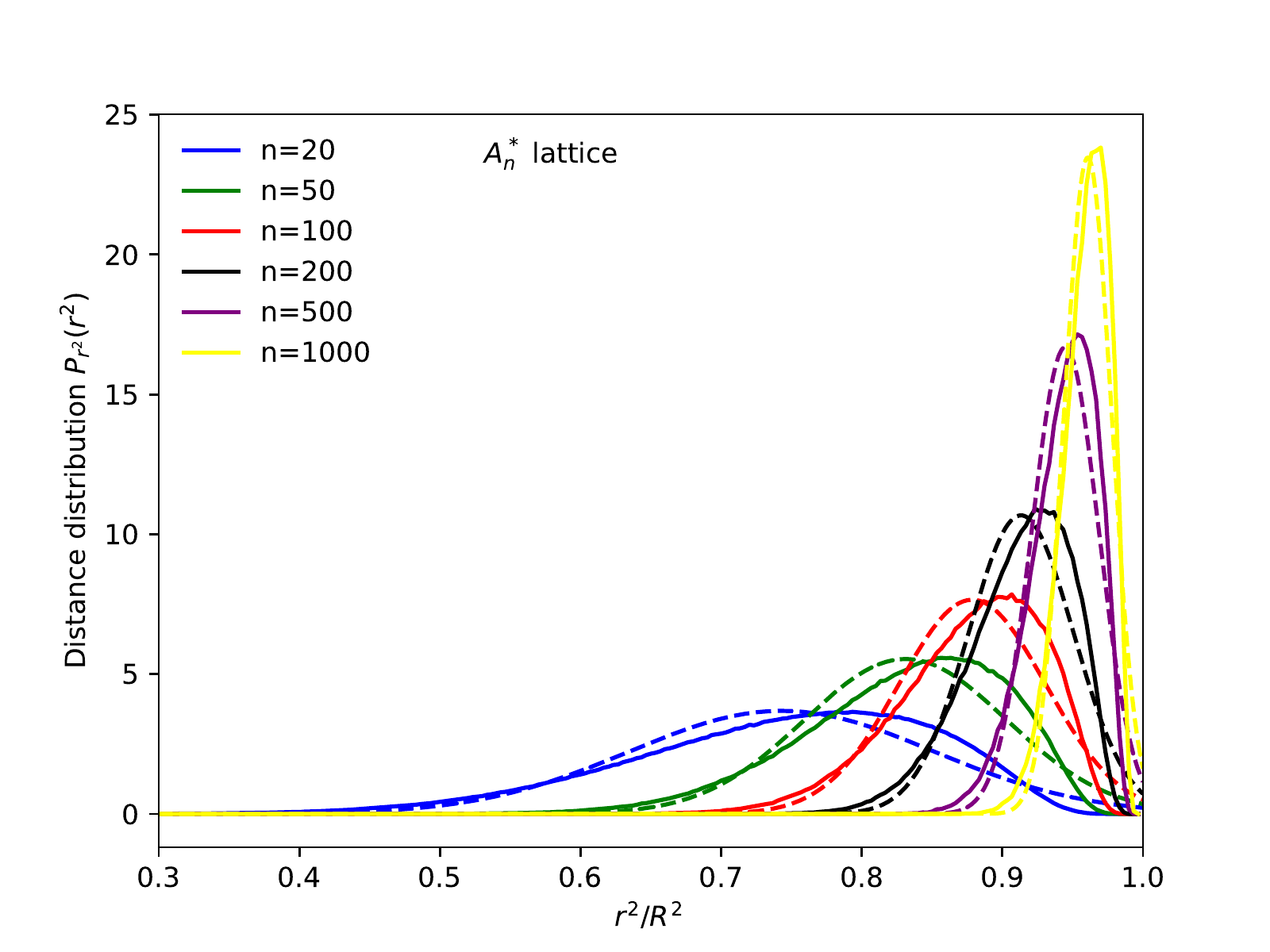}\non
    \ea
    \caption{\label{FigExactApprox} The probability distribution of
      the squared distance $P_{r^2}(r^2)$ is shown as the solid
      curves for the $\mathbb{Z}^n$ (left plots) and $A_n^*$ (right
      plots) lattices, for a varying number of dimensions $n$.  The top curves show small
      numbers of dimensions $n=2,4,6,8,10$ and the lower curves show
      larger numbers of dimensions $n=20,50,100,200,500$. The
      dotted curves show the Gaussian approximation given in
      \eq{e:gaussapprox}, with the correct mean and variance for
      the given lattice and dimension.   One can see
      that for the $\mathbb{Z}^n$ lattice, as expected from the
      central limit theorem, the Gaussian approximation approaches the
      true distribution as $n \to \infty$, which is a Dirac
      delta function peaked at $r^2/R^2 = 1/3$. For $A_n^*$, the
      central limit theorem does not apply, and the Gaussian
      approximation does not approach the true distribution for large
      dimension.  Nevertheless, as $n \to \infty$, the distributions
      approaches a Dirac delta function peaked at $r^2/R^2 =
      1$.}
  \end{center}
\end{figure*}

To compute the radius distribution function $P_{r^2}(r^2)$ for large
$n$, we make use of the central limit theorem \cite{mathews1970mathematical}.
Consider the distance \eq{2.4}.  In the large-$n$ limit it is the sum of
many independent random variables, each of which has the same
distribution.  Thus, we expect that it should approach a normal or
Gaussian distribution, characterized entirely by the mean and variance
of the distribution. 

We have already calculated the moments of $r^2$ for the
$\mathbb{Z}^n$ lattice.  The mean and variance are given by
\be\n{13}
\langle r^2 \rangle = \frac{1}{12} n \ell^2  = \frac{1}{3}R^2,
\ee
and 
\be\n{14}
\sigma^2 = \langle r^4 \rangle - \langle r^2 \rangle^2
= \frac{1}{180} n \ell^4
=  \frac{4}{45 n} R^4\, .
\ee
From these, the large-$n$ limit follows immediately.  Note that
as $n$ gets large, the variance vanishes, which means that the
distribution becomes sharply peaked.

If $n$ is large enough that the central limit theorem applies, then
the distribution of squared distance is a Gaussian normal distribution
\be\n{e:gaussapprox}
P_{r^2}(r^2) dr^2 = (2 \pi \sigma^2)^{-1/2} {\rm
  e}^{-(r^2 -R^2/3)^2/2\sigma^2} dr^2 .
\ee
Note that if the dimension $n$ is large, then this has vanishing support for
negative $r^2$, otherwise the normalization may be suitably adjusted.

In the $n \to \infty$ limit with fixed mismatch, the variance vanishes, and the
distribution approaches a Dirac delta function
\be\n{e:deltafnlimit} \lim_{n \to \infty} P_{r^2}(r^2) = \delta\left(r^2 -
\frac{1}{3}R^2\right)\,.
\ee
In Fig.~\ref{FigExactApprox} we show how this
limit is approached.  When $n$ is larger than 2, one has $2 \sigma^2 <
(R^2/3)^2$ and as soon as $n$ is a few times larger than this, the
Gaussian distribution becomes a good approximation to the actual
mismatch.

\subsection{$r^2$-distribution for the $A_n^*$ lattice}

The case of the $A_n^*$ lattice is not as simple.  The squared
distance is still a quadratic form which can be diagonalized, but the
variables which make it up are no longer independent, because they are
constrained by the boundaries of the WS.  It is unlike the
$\mathbb{Z}^n$ lattice, where these constraints are independent for
each variable.  Hence, the central limit theorem cannot be applied.

\begin{figure}
  \begin{center}
   \includegraphics[width=8.0cm]{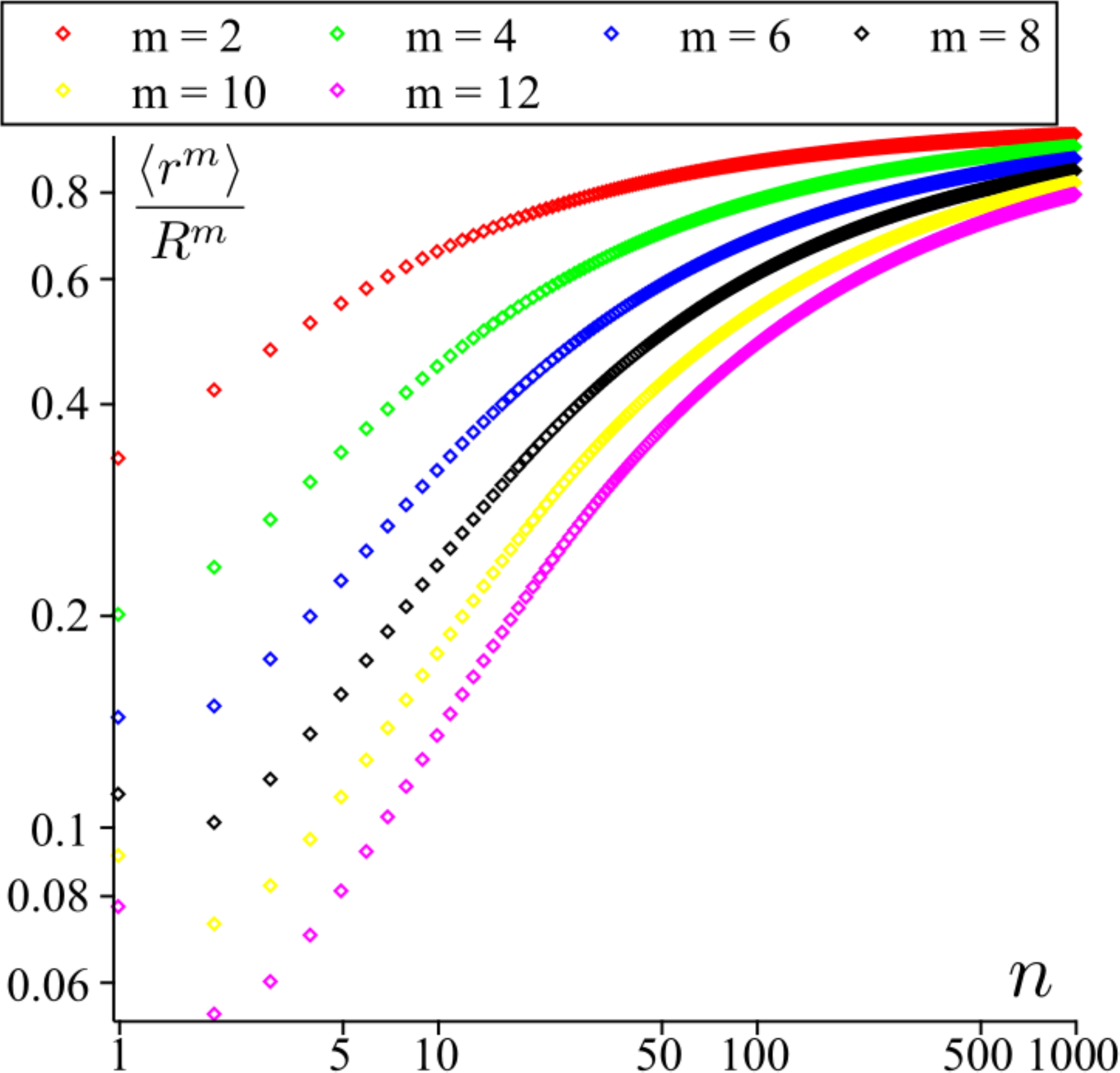}
   \caption{\label{Moments} The even moments $\langle r^{m}\rangle$ for $m=2,4,6,8,10,12$ in the units of $R^{m}$ for the $A_n^*$ lattice for dimensions from 1 to 1000 are shown in the ascending order.}
  \end{center}
\end{figure}

It is informative to examine the moments of $r^2$
defined by \eq{3.2}, which are computed exactly via recursion in 
Appendix~\ref{s:AnStarMoments}.
Fig.~\ref{f:analyticAstar} shows
the mean and variance of $r^2$ for the $A_n^*$ lattice.  One
immediately sees a significant difference when compared with the
$\mathbb{Z}^n$ lattice: at large dimension, the mean value of squared
radius $\langle r^2 \rangle$ approaches the squared WS radius $R^2$, whereas
for $\mathbb{Z}^n$ it is $1/3$ of that value.  As with the cubic
lattice, the variance approaches zero at large dimension, indicating
that the distribution is becoming sharply peaked. 
Some higher moments $\langle r^{2m} \rangle$ are shown in Fig.~\ref{Moments}: for large $n$ they asymptote to
$R^{2m}$.

\begin{figure}
  \begin{center}
\hspace{-0.8cm}\includegraphics[width=9.5cm]{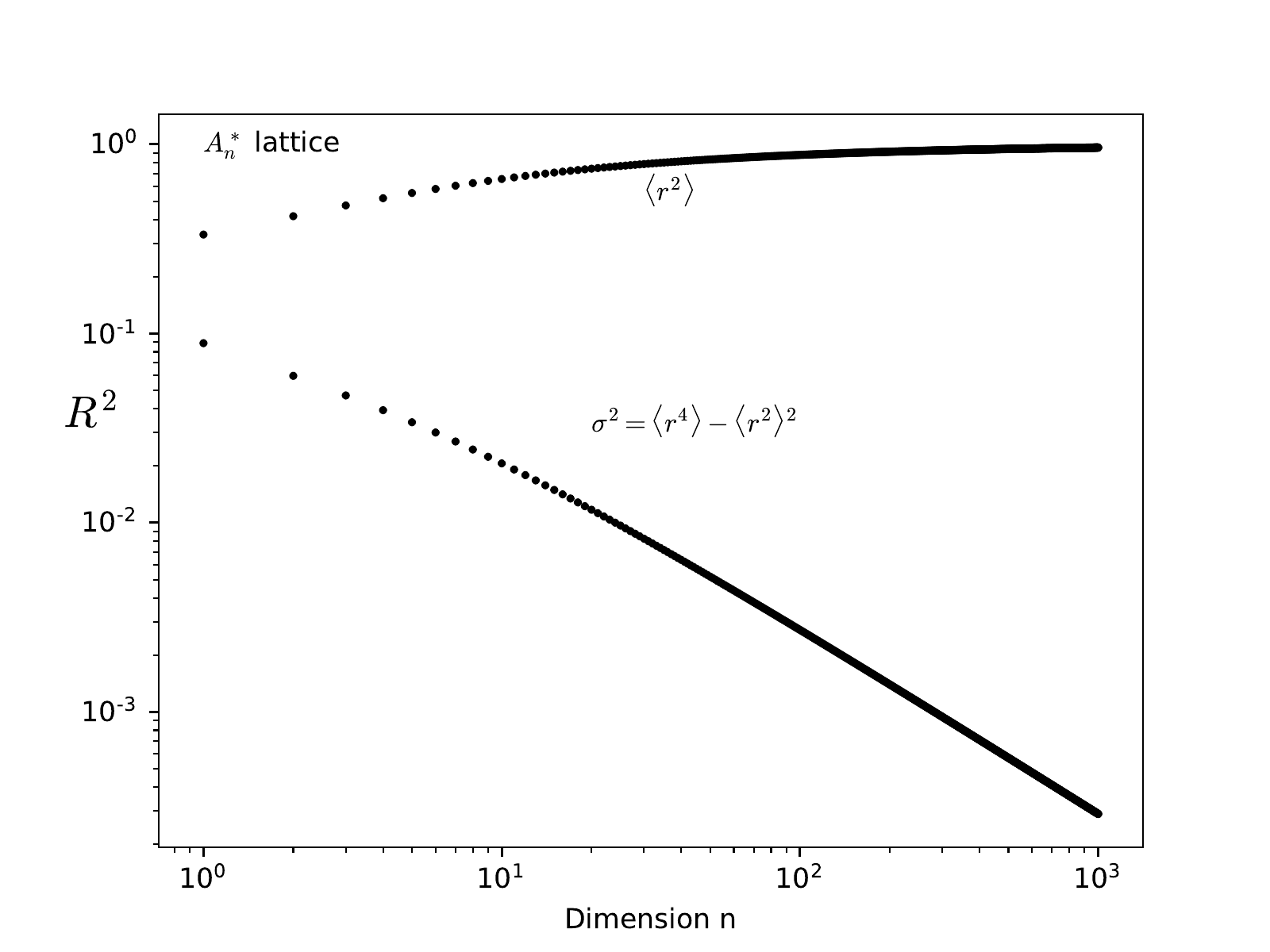}
   \caption{\label{f:analyticAstar} The mean and variance of the
     squared radius for the $A_n^*$ lattice for dimensions from 1 to
     1000, obtained exactly using the recursion in Appendix
     \ref{s:AnStarMoments}.  At large dimension the distribution is a
     narrow peak at the squared WS radius $R^2$.}
  \end{center}
\end{figure}

It is straightforward to study the distribution function numerically.
First, select points at random from within the FP, by drawing the
lattice coordinates $y_1,\cdots, y_n$ from independent uniform
distribution in the range $[0, \ell]$.  Then identify the closest
lattice point to $x = y_i \BM{e}_i$ and calculate the distance between
the two.  We now describe how to identify this closest lattice point.
(An algorithm is given in \cite{Conway} for $A_n$ as well as the
correspondence with the dual lattice $A_n^*$, but we were unable to
implement it.)

It is straightforward to show that the closest lattice point to $x$
must be one of the vertices of the FP.  Since there are $2^n$ such
vertices, when $n$ is large, it's not computationally feasible to
check the distances to all of them.  However, it is trivial to show
that the distance to the closest lattice point is unchanged if we
permute the ordering of the lattice coordinates $y_i$.  So the first
step of simplification is to reorder the lattice coordinate values of
$y_i$ in increasing order.

We now prove the following.  If $0\le y_1 \le \cdots \le y_n
\le \ell$ are the lattice coordinates of a point in the FP, then the
closest FP vertex has coordinates of the form $(0, \cdots, 0, \ell,
\cdots, \ell)$, where there are $k$ zeros followed by $(n-k)$ $\ell$'s.
The proof is by contradiction.

Suppose that the closest vertex to the point with lattice coordinates $(y_1, \cdots, y_n)$ is a point
with lattice coordinates
$A=(0, \cdots, 0, \ell, 0, \ell, \cdots, \ell)$ and is at squared
distance $r_A^2$.  We use $y_R$ to denote
the lattice coordinate value at the position of the rightmost zero, and
$y_L$ to denote the value at the leftmost $\ell$.  Now, construct a
different lattice vertex B, by swapping the leftmost $\ell$ with the
$0$ just to its right, so that $B=(0, \cdots, 0, \ell, \ell, \cdots, \ell)$, and
denote its squared distance from $y$ by $r_B^2$.  The difference between the squared
distances is
\ba
\non
r_A^2 - r_B^2 & = &\left(1+\frac{1}{n}\right) \biggl ((y_L-\ell)^2 +
y_R^2 - (y_R-\ell)^2 - y_L^2) \biggr)\\
& = & 2 \ell\left(1+\frac{1}{n}\right)(y_R - y_L).
\ea
Since the coordinates are ordered so that $y_L < y_R$, it
follows that $r_A^2 - r_B^2 > 0$ and thus that $A$ is {\it not} the
closest lattice vertex to $y$.  The same argument shows that swapping
a leftmost $\ell$ with a $0$ anywhere to its right will always decrease the
distance.  The result follows by induction.

This makes it computationally straightforward to identify the closest
vertex to any point inside the FP.  First, sort the lattice coordinates in
increasing order.  Then, calculate the distances to the $n+1$
vertices with coordinates of the form $(0, \cdots, 0, \ell, \cdots,
\ell)$ and select the minimum.

We have used this method to find $P_{r^2}(r^2)$ numerically for the
$A_n^*$ lattice, for dimensions from $n=1$ to $n=1000$.  This is
plotted in Fig.~\ref{FigExactApprox}.  In comparison with the
cubic lattice $\mathbb{Z}^n$, two differences are immediately
apparent.  The first is that as the dimension $n$ increases, the
distribution increasingly becomes peaked around the WS radius $R$, and
the second is that the Gaussian approximation (with the correct mean
and variance) is not good, because it does not fall off fast enough as
$r \to R$.

\section{Conclusion}

In this paper, we have computed and compared the loss fractions of two
template grids.  The first is based on the simple cubic lattice
$\mathbb{Z}^n$, and the second is based on the root lattice $A_n^*$,
which is a generalization of the two-dimensional hexagonal lattice.
In particular, we extend the results of \cite{NewAllenPaper} to the
case of large mismatch, by exploiting the spherical approximation
\cite{AllenSpherical}.

The main result is rather clear, and visible in the upper part of
Fig.~\ref{f4}. The slight advantages offered by the $A_n^*$ lattice
at small mismatch {\it decrease} at larger mismatch.  This can be
easily understood from the distribution of the squared radius for
points randomly selected within the Wigner-Seitz (WS) cell.  As the
dimension $n$ of parameter space increases, this distribution becomes
an increasingly narrow peak centered closer and closer to the squared
WS radius.

We believe that this behavior may be general, and true for any lattice
in the limit as the dimension $n\to \infty$.  To state it precisely,
the distribution function for the squared radius becomes an
increasingly narrow peak, which is true if and only if
\be
\lim_{n\to\infty} \langle r^{2m} \rangle = \langle r^2 \rangle^m,
\ee
with the understanding that the WS radius $R$ is held fixed during the
limiting process.  We have tried to prove this using Jensen's
inequality \cite{Perlman}, but are not convinced that our argument is
correct.

The final messages for the data analyst are simple ones.  First, a
fairly effective template-based search can be constructed at mismatch
values that are shockingly high in the quadratic approximation
(quadratic mismatch exceeding unity!).  Second, if the goal is to
detect as many signals as possible at fixed computing cost, there is
little motivation for using template banks based on sophisticated
lattices such as $A_n^*$.  These offer only minimal benefit when
compared with the humble cubic lattice $\mathbb{Z}^n$, and that minor
advantage diminishes as the template separation increases.

\section{Acknowledgments}

We thank Mathieu Dutour Sikiri\'c for bringing the thinnest known
lattices of \cite{Sikiric_2008} to our attention.

\appendix

\section{\label{s:zlatticemoments} Even moments of the $\mathbb{Z}^n$ lattice}

For the $\mathbb{Z}^n$ lattice, the general even-order moment can be computed as follows.  One uses the multinomial expansion to write
\be
\n{e:generalterm}
\langle r^{2m}\rangle = \sum_{k_1+\cdots+k_n=m} \binom{m}{k_1,\cdots,k_n} \prod_{i=1}^n \langle x_i^{2k_i} \rangle,
\ee
where the sum is over all non-negative integer $k_i$ whose sum equals $m$. The multinomial coefficient is
\be
\binom{m}{k_1,k_2, \cdots,k_n} = \frac{m!}{k_1 ! k_2! \cdots k_n!},
\ee
and the coordinate moments are 
\be
\langle x^{2k} \rangle = \frac{1}{\ell}\int_{-\ell/2}^{\ell/2} x^{2k} dx = \frac{1}{2k+1}\left(\frac{\ell}{2}\right)^{2k}.
\ee
In the sum \eq{e:generalterm}, there are many identical terms on the
r.h.s. which are obtained by permutation of the indices of the $k_i$.
The number of these identical terms depends upon the number of
distinct non-zero values taken by the $k_i$, which in turn depends
upon the dimension $n$. 

Suppose that for each term, the non-zero
$k_i$ are sorted in increasing order; there at most $m$ of them. Let $q \le m$
denote the number of these non-zero $k_i$, and
let $n_1$ denote the number
of $k_i$ which have the smallest value, $n_2$ the next smallest, and
so on; the sum is bounded by $\sum_i n_i \le m$.  Then the number of
equivalent (under permutation) terms which appear on the r.h.s. of \eq{e:generalterm}
is
equal to the number of ways in which $n_1$ coordinates can be chosen
from the $n$, and $n_2$ can be chosen from the remaining $n-n_1$, and
so on.  This is
\ba
\non
&& N(k_1,\cdots, k_q) \\
\non
&=& \binom{n}{n_1} \binom{n-n_1}{n_2} \times \cdots \times \binom{n-n_1 - \cdots - n_{p-1}}{n_p}\\
&= & \frac{n!}{n_1! n_2! \cdots n_p! (n-n_1 - \cdots - n_p)! },
\ea
where the quantities in the second line are the standard binomial (choice) coefficients;
the r.h.s. is a polynomial in $n$ of order $\le m$.
Thus one obtains
\be
\n{e:simpleform}
\langle r^{2m}\rangle =   \hspace{-1.5em} \sum_{k_1+\cdots+k_q=m} \frac{\binom{m}{k_1,\cdots,k_q} N(k_1, \cdots, k_q)}{(2k_1+1) \cdots (2k_q+1)} \left(\frac{\ell}{2}\right)^{2m} \, ,
\ee
where the sum is over all distinct (under permutation) partitions
$k_i$.

For example, for $m=5$, the r.h.s. of \eq{e:simpleform} has
seven terms, with the following
sets of $k_i$: $\{1,1,1,1,1\}, \{1,1,1,2\}, \{1,2,2\}, \{1,1,3\},
\{2,3\}, \{1,4\}$, and $\{5\}$. Respectively, these have $n_i$ given
by $\{5\}, \{3,1\}, \{1,2\}, \{2,1\}, \{1,1\}, \{1,1\}$, and $\{1\}$,
with corresponding $N$ given by $n(n-1)(n-2)(n-3)(n-4)/5!$,
$n(n-1)(n-2)(n-3)/3!$, $n(n-1)(n-2)/2!$, $n(n-1)(n-2)/2!$, $n(n-1)$,
$n(n-1)$ and $n$.  Thus one obtains 
\ba
\non
&& \left( \frac{\ell}{2} \right)^{-10} \langle r^{10} \rangle   = \\ \non && 
\binom{5}{1,1,1,1,1} \frac{n(n-1)(n-2)(n-3)(n-4)}{5! \,  3^5} +\\ \non &&
\binom{5}{1,1,1,2}  \frac{n(n-1)(n-2)(n-3)}{3! \, 3^3 5} + \\ \non && 
\binom{5}{1,2,2} \frac{n(n-1)(n-2)}{2! \,  5^2 \, 3} +  
\binom{5}{1,1,3} \frac{n(n-1)(n-2)}{2! \, 3^2 \, 7} +  \\ \non &&
\binom{5}{2,3} \frac{n(n-1)}{ 5 \cdot 7} + 
\binom{5}{1, 4} \frac{n(n-1)}{ 3 \cdot 9} + 
\binom{5}{5} \frac{n}{ 11 }\, . 
\ea
This simplifies, to give the tenth moment of \eq{3.3}.

The supplementary materials for this manuscript include a short
Mathematica script to calculate arbitrary even moments of the
$\mathbb{Z}^n$ lattice.

\section{\label{s:AnStarMoments} Even moments of the $A_n^*$  lattice}

Here we give a general expression for computation of any even moment of the $A_n^*$ lattice. The computation technique is a
generalization of Chapter 21 Section 3.F of \cite{Conway}, where it is
used to find the second moment.

\begin{figure}[htb]
  \begin{center}
   \includegraphics[width=5.0cm]{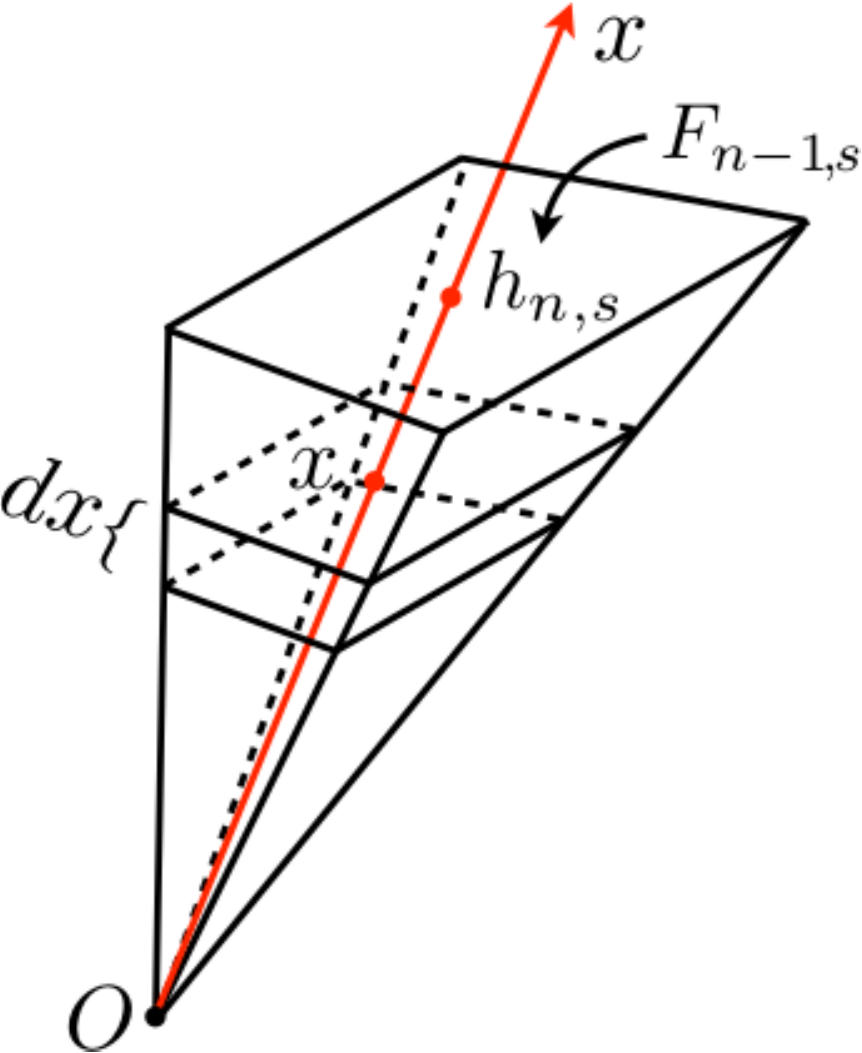}
   \caption{\label{pyramid} The $n$-dimensional pyramid associated with $(n-1)$-dimensional face $F_{n-1,s}$. The point $O$ is such that for all congruent faces $F_{n-1,s}$ the associated pyramids are congruent. The axis $Ox$ is perpendicular to the face $F_{n-1,s}$, and $h_{n-1,s}$ is the distance form $O$ to the face. The increment $dx$ is thickness of the slab at $x$.}
  \end{center}
\end{figure}

The un-normalized and normalized $p$'th moments of a region or object $D$ are defined as
\be\n{e:Um} 
U_p(D) = \int_{D} r^p dV, \text{ and } I_p(D) = U_p(D) /U_0(D),
\ee
where $D$ is the domain of integration and the radius $r$ is measured from the origin $O$ (see Fig.~\ref{pyramid}).

The WS cell in dimension $n$ is called a permutohedron and is denoted
$P_n$.  It has a complex shape with $(n+1)!$ vertices and $2^{n+1}-2$
faces. According to the definition \eq{e:Um}, $U_0(P_n)$ is the volume of the
WS cell $P_n$.  The normalized $m$'th moment $I_p(P_n)$ is
obtained by dividing out the volume.

Note that the length conventions used in this Appendix follow \cite{Conway}, and
differ from the conventions used in the remainder of this paper.  To
transform a quantity associated with $P_n$ with dimensions of length${}^d$ in this Section
into the units used in the remainder of the paper, multiply by
\be
\label{e:conventions}
\biggl[\frac{\ell^2}{n(n+1)} \biggr]^{d/2}.
\ee
For example, in the conventions of this Section, the point in $P_n$ most distant
from the center has
squared
radius $n(n+1)(n+2)/12$, which should be compared
with \eq{2.15}, and the volume is
$U_0(P_n) = (n+1)^{n-1/2}$, which should be compared with \eq{2.11a}.

Each face of $P_n$ is the direct product of a pair of
lower-dimensional permutohedrons \footnote{A face of $P_n$ is only the
  direct product (metrically as well as geometrically!) of
  lower-dimensional faces of $P_n$ if we follow the the
  ``dimension-dependent'' length conventions of \cite{Conway}.}.  For
$n$ even there are $n/2$ types of faces and for $n$ odd there are
$(n+1)/2$ types of faces.  Following \cite{Conway} the different types
of faces are labeled by $s=0, \cdots, n-1$.  A face of type $s$, $F_{n-1,s}$, is the
Cartesian product, $F_{n-1,s}=P_s \times P_{n-s-1}$; faces of type $s$ and
faces of type $n-s-1$ are equivalent.  The number of faces of type $s$
is the binomial coefficient
\be
{{n+1}\choose{s+1}}.
\ee
The squared distance from the center of $P_n$ to the center of
a face of type $s$ is
\be
h^2_{n,s} = \frac{1}{4} (s+1)(n-s)(n+1).
\ee
By symmetry, the line from the center of $P_n$ to the center of any
face is orthogonal to the face.  We call this the center line to the face.

Because the faces are formed from lower-dimensional permutohedrons,
the moments may be calculated by recursion.  We divide $P_n$ into
generalized pyramids, one face at a time, by taking the bundle of all
line segments that begin at the center of the $P_n$ and extend to
anywhere in that face.  These pyramids are disjoint (apart from a set
of measure zero on their boundaries) and their union is $P_n$.  To
compute the moments of $P_n$, we compute the moments of the pyramids
and sum them.

The $m$'th moment of each pyramid can be found with elementary
calculus.  We slice each pyramid into slabs of thickness $dx$, where
$x\in[0,h_{n,s}]$ is a a coordinate that runs along the center line to a
face of type $s$, and the slicing
is orthogonal to the center line shown in Fig.~\ref{pyramid}.
Each slab has $n$-volume
\be\n{svolume}
dV = \frac{x^{n-1}}{h_{n,s}^{n-1}} U_0(P_{s})U_0(P_{n-s-1}) dx,
\ee
so by integration over $x$ the volume of the pyramid is
\be
\int^{h_{n,s}}_{0} dV  = \frac{1}{n} h_{n,s}   U_0(P_{s})U_0(P_{n-s-1}).
\ee
Summing over all faces gives
\be
U_0(P_n) = {1 \over n} \sum_{s=0}^{n-1} {{n+1}\choose{s+1}} h_{n,s}  U_0(P_{s})U_0(P_{n-s-1}).
\ee
This recursion relation, together with the initial value $U_0(P_0)=1$, determines the volume $U_0(P_n)$ for dimensions $n>0$.

To construct a general recursion relation for an arbitrary even moment $U_{2m}(P_n)$, $m=0,1,2,3,...$, we begin with the expression
\be\n{e:U2m}
U_{2m}(P_n) = \sum_{s=0}^{n-1} {{n+1}\choose{s+1}} U_{2m}(P_{n,s})\,,
\ee
where $U_{2m}(P_{n,s})$'s are the moments  of $n$-dimensional pyramids $P_{n,s}$ into which a permutohedron $P_n$ is decomposed. Every such moment can be calculated by using the definition \eq{e:Um}, substituting for $r^2$ the expression
\be
r^2=\frac{x^2}{h^2_{n,s}}\rho^2_{n-1,s}+x^2\,,
\ee
where $\rho_{n-1,s}$ is the distance from the point of intersection of the axis $Ox$ with the face $F_{n-1,s}$ to arbitrary point of the face (see Fig.~\ref{pyramid}), as follows:
\ba
U_{2m}(P_{n,s})&=&\int_{F_{n-1,s}} dV \, \int^{h_{n,s}}_0 dx \, \left(\frac{x^2}{h^2_{n,s}}\rho^2_{n-1,s}+x^2\right)^m\non\\
&=&\sum_{k=0}^{m}{{m}\choose{k}}\int_{F_{n-1,s}}\hspace{-0.5cm}\rho^{2k}_{n-1,s}dV\int^{h_{n,s}}_0\frac{x^{2m+n-1}}{h_{n,s}^{n-1+2k}}dx\,,\non
\ea
where the volume element $dV$ is in the face $F_{n-1,s}$. (For odd moments $m=k+1/2$, where $k=0,1,2,...$ the finite sum in the expression above is replaced by an infinite series.) Using the definition \eq{e:Um} for the moments $U_{2k}(F_{n-1,s})$ of faces $F_{n-1,s}$ (here, with origin at the center of the face) and integrating over $x$ we obtain
\be
U_{2m}(P_{n,s})=\frac{h_{n,s}^{2(m-k)+1}}{n+2m}\sum_{k=0}^{m}{{m}\choose{k}}U_{2k}(F_{n-1,s})\,.
\ee
Substituting this expression into \eq{e:U2m} we obtain
\ba\n{2m}
U_{2m}(P_n)&=&{1 \over n+2m} \sum_{s=0}^{n-1}\sum_{k=0}^{m}{{n+1}\choose{s+1}}{{m}\choose{k}}\non\\
&\times&h_{n,s}^{2(m-k)+1}U_{2k}(F_{n-1,s})\,.
\ea
The next step is to consider the face $F_{n-1,s}$ as the Cartesian product $P_s \times P_{n-s-1}$ and apply again the definition \eq{e:Um} to the moments $U_{2k}(F_{n-1,s})$, with the origin at the center of the face.  Replace $r^2$ with 
\be
\label{e:rhodef}
\rho^2_{n-1,s}=\rho^2_{s}+\rho^2_{n-1-s}\,,
\ee
and use the binomial theorem to raise \eq{e:rhodef} to power $k$.
Employing the definition \eq{e:Um} for the moments $U_{2j}(P_{s})$ and $U_{2(k-j)}(P_{n-1-s})$,
we obtain
\be\n{2k}
U_{2k}(F_{n-1,s})=\sum_{j=0}^{k}{{k}\choose{j}}U_{2j}(P_{s})U_{2(k-j)}(P_{n-s-1})\,.
\ee
Finally, substituting \eq{2k} into \eq{2m}, we obtain the following relation for the even moments of $P_n$:
\ba\n{mrec}
U_{2m}(P_n) &=& {1 \over n+2m} \sum_{s=0}^{n-1}\sum_{k=0}^{m}\sum_{j=0}^{k} {{n+1}\choose{s+1}} {{m}\choose{k}}{{k}\choose{j}}\non\\
&\times&h^{2(m-k)+1}_{n,s}U_{2j}(P_{s})U_{2(k-j)}(P_{n-s-1})\,.
\ea
This recursion relation, together with the initial values
$U_{0}(P_0)=1$ and $U_{2m}(P_0)=0$, for $m=1,2,3,...$, defines an
arbitrary even-order moment.

In Tables~\ref{t:MomentsAnStar} and \ref{t:ExactAstarMoments} we give
numerical and exact values for the even moments $\langle r^{2m}
\rangle$ obtained from $U_{2m}(P_n)$, for dimensions $n < 16$.  The
un-normalized moments are computed using the recursion relation
\eq{mrec}. The normalized moments $I_{2m}(P_n)$ are then defined by
\ref{e:Um}.  Both of these follow the conventions of Conway and
Sloane, Chapter 21 Section 3F \cite{Conway}.  They are then re-scaled
following \eq{e:conventions} with $d=2m$ to obtain $\langle r^{2m}
\rangle$, which are in the conventions used everywhere else in this
paper.

The following lines of Mathematica are sufficient to calculate the
arbitrary even moments $U_{2m}(P_n)={\rm U[m,n]}$ up to dimensions of
several thousand.  The normalized moments are $I_{2m}(P_n)={\rm II[m,n]}$,
and the moments (with the length conventions used in the remainder of
this paper, as they appear in Tables~\ref{t:MomentsAnStar} and
\ref{t:ExactAstarMoments}) are $<r^{2m}> = {\rm Mo[m,n]}$.
\newpage
\begin{Verbatim}[frame=single] 
h[n_,s_]:= h[n,s] = Sqrt[(s+1)(n-s)(n+1)]/2
U[m_,0 ]:= U[m,0] = If[m==0, 1, 0]
U[m_,n_]:= U[m,n] = Simplify[Sum[
 Binomial[n+1,s+1] Binomial[m,k] Binomial[k,j] 
 h[n,s]^(2 m - 2 k + 1) U[j, s] U[k-j, n-s-1], 
{s, 0, n-1},{k, 0, m},{j, 0, k}]/(n+2 m)]
II[m_,n_]:=U[m,n]/U[0,n]
Mo[m_,n_]:=II[m,n](12 R^2/(n(n+1)(n+2)))^m
\end{Verbatim}
\twocolumngrid
\begin{Verbatim}








\end{Verbatim}

\begin{table*}[htb]
  \begin{tabular}{|*{1}{*{7}{|l}|}|}
$n$ & $\langle r^2 \rangle /R^2$ & $\langle r^4 \rangle /R^4 $ & $\langle r^6 \rangle/ R^6$ & $\langle r^8 \rangle /R^8$ & $\langle r^{10} \rangle /R^{10}$ & $\langle r^{12} \rangle /R^{12}$ \\ \hline
$1$ & $0.3333$ & $0.2000$ & $0.1429$ & $0.1111$ & $0.0909$ & $0.0769$ \\
$2$ & $0.4167$ & $0.2333$ & $0.1482$ & $0.1013$ & $0.0727$ & $0.0541$ \\
$3$ & $0.4750$ & $0.2727$ & $0.1728$ & $0.1167$ & $0.0824$ & $0.0602$ \\
$4$ & $0.5187$ & $0.3083$ & $0.1987$ & $0.1353$ & $0.0959$ & $0.0701$ \\
$5$ & $0.5529$ & $0.3397$ & $0.2233$ & $0.1540$ & $0.1101$ & $0.0810$ \\
$6$ & $0.5807$ & $0.3672$ & $0.2460$ & $0.1720$ & $0.1242$ & $0.0921$ \\
$7$ & $0.6038$ & $0.3914$ & $0.2669$ & $0.1891$ & $0.1380$ & $0.1031$ \\
$8$ & $0.6234$ & $0.4130$ & $0.2862$ & $0.2052$ & $0.1512$ & $0.1139$ \\
$9$ & $0.6404$ & $0.4324$ & $0.3039$ & $0.2204$ & $0.1639$ & $0.1244$ \\
$10$ & $0.6552$ & $0.4499$ & $0.3202$ & $0.2346$ & $0.1760$ & $0.1346$ \\
$11$ & $0.6683$ & $0.4657$ & $0.3354$ & $0.2481$ & $0.1876$ & $0.1444$ \\
$12$ & $0.6800$ & $0.4802$ & $0.3496$ & $0.2608$ & $0.1986$ & $0.1539$ \\
$13$ & $0.6905$ & $0.4936$ & $0.3628$ & $0.2728$ & $0.2092$ & $0.1630$ \\
$14$ & $0.7001$ & $0.5058$ & $0.3751$ & $0.2842$ & $0.2193$ & $0.1719$ \\
$15$ & $0.7088$ & $0.5172$ & $0.3867$ & $0.2950$ & $0.2290$ & $0.1804$ \\
\end{tabular}

  \caption{\label{t:MomentsAnStar} Numerical values for the first six
    even moments of the WS cell $P_n$ of the $A_n^*$ lattice
    for dimensions $n=1,\dots, 15$, where the covering radius $R$ is
    given in \eq{2.15}. The exact values of these moments are given in
    Table~\ref{t:ExactAstarMoments}.  In the text we argue that
    $\lim_{n\to\infty} \langle r^2 \rangle/R^2 =1$, and that
    $\lim_{n\to\infty} \langle r^2m \rangle = \langle r^2
    \rangle^m$. Hence, as $n \to \infty$, all of these table entries should
    approach unity.  }
\end{table*}

\begin{turnpage}
  \begin{table*}
      \caption{\label{t:ExactAstarMoments} Exact values of the
        lowest-order even moments for the $A_n^*$ lattice, as given by
        the recursion relationship derived in
        Appendix~\ref{s:AnStarMoments}. Numerical values may be found
        in Table~\ref{t:MomentsAnStar}.}
      \begin{tabular}{|*{1}{*{7}{|l}|}|}
$n$ & $\langle r^2 \rangle /R^2$ & $\langle r^4 \rangle /R^4$ & $\langle r^6 \rangle /R^6$ & $\langle r^8 \rangle /R^8$ & $\langle r^{10} \rangle / R^{10}$ & $\langle r^{12} \rangle / R^{12}$ \\ \hline
&&&&&&\\
$1$ & $\frac{1}{3}$ & $\frac{1}{5}$ & $\frac{1}{7}$ & $\frac{1}{9}$ & $\frac{1}{11}$ & $\frac{1}{13}$ \\ 
&&&&&&\\
$2$ & $\frac{5}{12}$ & $\frac{7}{30}$ & $\frac{83}{560}$ & $\frac{319}{3150}$ & $\frac{403}{5544}$ & $\frac{1517}{28028}$ \\
&&&&&&\\
$3$ & $\frac{19}{40}$ & $\frac{409}{1500}$ & $\frac{27217}{157500}$ & $\frac{38281}{328125}$ & $\frac{1427333}{17325000}$ & $\frac{539235971}{8957812500}$ \\
&&&&&&\\
$4$ & $\frac{389}{750}$ & $\frac{37}{120}$ & $\frac{312953}{1575000}$ & $\frac{15977293}{118125000}$ & $\frac{41514229}{433125000}$ & $\frac{10041681373}{143325000000}$ \\
&&&&&&\\
$5$ & $\frac{209}{378}$ & $\frac{22469}{66150}$ & $\frac{1085536}{4862025}$ & $\frac{131002604}{850854375}$ & $\frac{43271601607}{393094721250}$ & $\frac{120676030448681}{1490484151406250}$ \\
&&&&&&\\
$6$ & $\frac{78077}{134456}$ & $\frac{423151}{1152480}$ & $\frac{66683227}{271063296}$ & $\frac{2855332489}{16602626880}$ & $\frac{3049146605765}{24545323579392}$ & $\frac{561024599680723}{6091703033794560}$ \\
&&&&&&\\
$7$ & $\frac{89035}{147456}$ & $\frac{1515163}{3870720}$ & $\frac{3645004793}{13655900160}$ & $\frac{813284979379}{4301608550400}$ & $\frac{43871720179117}{317974904045568}$ & $\frac{47003442367289761}{455737531223310336}$ \\
&&&&&&\\
$8$ & $\frac{3313213}{5314410}$ & $\frac{146336671}{354294000}$ & $\frac{127738979843}{446410440000}$ & $\frac{23552471440009}{114791256000000}$ & $\frac{343658283760349}{2272866868800000}$ & $\frac{26512131817729691533}{232684745693400000000}$ \\
&&&&&&\\
$9$ & $\frac{1100657}{1718750}$ & $\frac{117719}{272250}$ & $\frac{895784540747}{2947957031250}$ & $\frac{535892016104111}{2432064550781250}$ & $\frac{217007542831819286}{1324259147900390625}$ & $\frac{7422858954915606122659}{59651253317173095703125}$ \\
&&&&&&\\
$10$ & $\frac{18539106101}{28295372292}$ & $\frac{9643229133}{21435888100}$ &  $\frac{396436661186}{1237922537775}$ & $\frac{7667873302060633}{32681154997260000}$ & $\frac{556729110272163449}{3163535803734768000}$ & $\frac{4395579391379571244226791}{32656586938990809795000000}$ \\
&&&&&&\\
$11$ & $\frac{25941989}{38817792}$ & $\frac{6463162733}{13877360640}$ & $\frac{1863816574691}{5556495200256}$ & $\frac{1400046042361}{5643315437760}$ & $\frac{1406514743813641321}{7499234843102306304}$ & $\frac{26425247449515274378841}{182976643149919334876160}$ \\
&&&&&&\\
$12$ & $\frac{17061288862565}{25090245516518}$ & $\frac{389281868476643}{810607932072120}$ & $\frac{135373127776116799}{387267939547455330}$ & $\frac{3186173305108287135529}{12217012599590390810400}$ & $\frac{194307923663053340412317}{978338368975198496096832}$ & $\frac{44840145568064414926573819}{291366954251159115746656512}$ \\
&&&&&&\\
$13$ & $\frac{81923134171}{118639604580}$ & $\frac{2718612928907}{5508267355500}$ & $\frac{4772968402612073}{13157873645450625}$ & $\frac{26384102217284296859}{96710371294062093750}$ & $\frac{8775008810370829285906}{41949735388321000865625}$ & $\frac{3034165287365868825626600411}{18609951361643904009012890625}$ \\
&&&&&&\\
$14$ & $\frac{3588341852791}{5125781250000}$ & $\frac{5807913344549}{11481750000000}$ & $\frac{6330843027118853801}{16878172500000000000}$ & $\frac{3357739197420742084967}{11814720750000000000000}$ & $\frac{9575820529342227197910377}{43667207892000000000000000}$ & $\frac{17383801168332194619594087887}{101149132462560000000000000000}$ \\
&&&&&&\\
$15$ & $\frac{423929152911307}{598134325510144}$ & $\frac{49305729309609649}{95327658128179200}$ & $\frac{210551088639050555933}{544511583228159590400}$ & $\frac{19505342217281370335341}{66119263677705093120000}$ & $\frac{271786289285385309816213443}{1186973021542161831690240000}$ & $\frac{664211366929553172581639942783}{3682314032738752046039040000000}$ \\
\end{tabular}

\end{table*}
\end{turnpage}

\end{document}